\pdfoutput=1 
\documentclass[fleqn,usenatbib]{mnras}
\usepackage{amsmath}
\usepackage{newtxtext,newtxmath}
\usepackage[T1]{fontenc}
\usepackage{ae,aecompl}
\usepackage{colortbl}
\usepackage{graphicx}
\usepackage{pdflscape}

\newcommand{\bs}{\boldsymbol}

\usepackage{graphicx}	% Including figure files

\title[Impact of thermal effects on eccentricity and
inclination]{Impact of thermal effects on the evolution of
  eccentricity and inclination of low-mass planets}

\author[S. Fromenteau \& F. Masset]{
S\'ebastien Fromenteau\thanks{sfroment@icf.unam.mx}
and Fr\'ed\'eric S. Masset
\\
Instituto de Ciencias F\'isicas, Universidad Nacional Aut\'onoma de M\'exico, Av. Universidad s/n, 62210 Cuernavaca, Mor., Mexico}

\date{Accepted XXX. Received YYY; in original form ZZZ}

\pubyear{2019}

\begin{document}
\label{firstpage}
\pagerange{\pageref{firstpage}--\pageref{lastpage}}
\maketitle

\begin{abstract}
  Using linear perturbation theory, we evaluate the time-dependent
  force exerted on an eccentric and inclined low-mass planet embedded
  in a gaseous protoplanetary disc with finite thermal diffusivity
  $\chi$. We assume the eccentricity and inclination to be small
  compared to the size of the thermal lobes
  $\lambda\sim(\chi/\Omega)^{1/2}$, itself generally much smaller than
  the scalelength of pressure $H$. When the planet is non-luminous, we
  find that its eccentricity and inclination are vigorously damped by
  the disc, over a timescale shorter by a factor $H/\lambda$ than the
  damping timescale in adiabatic discs. On the contrary, when the
  luminosity-to-mass ratio of the planet exceeds a threshold that
  depends on the disc's properties, its eccentricity and inclination
  undergo an exponential growth. In the limit of a large luminosity,
  the growth rate of the eccentricity is 2.5~times larger than that of
  the inclination, in agreement with previous numerical
  work. Depending on their luminosity, planetary embryos therefore
  exhibit much more diverse behaviours than the mild damping of
  eccentricity and inclination considered hitherto.
\end{abstract}


%%%%%%%%%%%%%%%%% BODY OF PAPER %%%%%%%%%%%%%%%%%%

\section{Introduction}
\label{sec:introduction}
Most analytic studies of planet-disc interactions have long been
limited to the barotropic approximation. The inclusion of
thermodynamics was introduced more than a decade ago, but was mainly
restricted to non-linear effects tackled through numerical
simulations, with a focus either on giant planets \citep{gda2003} or
on the non-linear dynamics of the corotation torque \citep{pm06}.  The
inclusion of non-barotropic effects in a linear analysis has been
worked out by \citet{2014ApJ...782..112T}, who performed a study of
the corotation torque in a non-barotropic disc. Studies of the role of
thermal diffusion itself, however, was even longer restricted to
non-linear effects, especially in the modelling of the saturation of
the corotation torque of intermediate-mass planets (above a few Earth
masses), either through numerical simulations or toy models of the
horseshoe dynamics
\citep{2010ApJ...723.1393M,pbk11,2017arXiv170708988J}.  The impact of
thermal diffusion on the interaction between the disc and low-mass
planets using linear perturbation theory has been studied more
recently. It was firstly noticed in numerical simulations of
non-luminous\footnote{Throughout this work we use this expression for
  planets that do not inject heat into the surrounding gas.}, low-mass
planets embedded in radiative\footnote{Discs in which thermal
  diffusion is effected by radiative transfer.} discs by
\citet{2014MNRAS.440..683L}, who found that the torque on a low-mass
planet on a circular orbit can be significantly more negative when
thermal diffusion is included than when it is not. This effect, dubbed
\emph{cold finger} by these authors, was later described by
\citet{2017MNRAS.472.4204M} who provided an analytic expression for
the corresponding torque component. Little is known, however, on the
impact of thermal diffusion on the gravitational interaction between
an eccentric or inclined non-luminous planet and a gaseous
disc. \citet{2017arXiv170401931E} performed numerical experiments in
isothermal and radiative discs, and found the eccentricity and
inclination damping to be much stronger in the latter than in the
former, but this effect, which was not the primary focus of that work,
was not systematically quantified, and may have been misrepresented by
the low resolution with which it was captured.

While the release of heat into the ambient gas by accreting planets
has been studied for more than two decades
\citep[e.g.][]{1996Icar..124...62P}, such studies considered the
planet to be at rest in a uniform medium in order to use
1D grids, and the feedback of heat release on the
planetary orbit was ignored. It was until recently that the heat
release was incorporated, in a highly simplified manner, to numerical
simulations of planet-disc interactions.  \citet{2015Natur.520...63B}
have found that luminous planetary embryos in the Earth-mass range may
undergo outwards migration if their luminosity is above a threshold
that should be easily overcome if they are subjected to fast pebble
accretion. The effect at the origin of this outward migration shares
many similarities with the ``cold finger'' effect of
\citet{2014MNRAS.440..683L}, to the point that both effects can be
unified into a single description
\citep{2017MNRAS.472.4204M}. \citet{2017arXiv170401931E} performed a
follow-up study of the work of \citet{2015Natur.520...63B}, by
relaxing the constraint of a circular and coplanar orbit. They found
that planetary embryos, if sufficiently luminous, undergo a growth of
eccentricity and inclination. A similar result holds in
2D calculations: \citet{2017arXiv170606329C} found that
luminous embryos embedded in 2D discs experience an
eccentricity growth. They called this effect the \emph{hot trail
  effect}. The disc's response to the heat release indeed adopts a
trailing, cometary shape for eccentricities well below the disc's
aspect ratio \citep{2017arXiv170401931E}. Unlike the response carried
by density waves, the effect of heat release can therefore be captured
by a simple calculation of dynamical friction even in the subsonic
regime \citep{2017MNRAS.465.3175M}. Note however that, again, the net
force arising from thermal effects in the case of dynamical friction
is positive (\emph{i.e.}, a thrust) if the planet is
\emph{sufficiently} luminous \citep{2019MNRAS.483.4383V}.

From the above we see that thermal effects on low-mass planets have
been studied both using linear perturbation theory and through
numerical simulations for planets in circular orbits and for
perturbers in unsheared, homogeneous media, adequate for the
description of planets with a sizeable eccentricity or
inclination. However, there has not been any analytical study of the
regime of small eccentricities and inclinations, when the disc
response is not that of a simple ``cometary'' trail captured by a
dynamical friction calculation. The purpose of this work is to
provide this missing part, so as to give analytic expressions for the
excitation or damping of the eccentricity and inclination of a
low-mass planet that can be used in future models of planetary
formation, and to shed some light on the different behaviours observed
so far in numerical experiments. We present our governing equations in
section~\ref{sec:governing-equations} and work out the density
response in section~\ref{sec:density-response}. The force arising from
thermal effects is then worked out in
section~\ref{sec:force-expression}, and the time evolution of the
planet's orbital parameters is derived in
section~\ref{sec:eccentr-incl-growth}. We finally discuss our results
in section~\ref{sec:discussion} and draw our conclusions in
section~\ref{sec:conclusion}.

\section{Governing equations}
\label{sec:governing-equations}
We consider a planet of mass $M$ embedded in a protoplanetary gaseous
disc on a prograde, slightly eccentric and slightly inclined
orbit. The central star has a mass $M_{\star}$, the disc has a surface
density $\Sigma(r)$ and an angular velocity $\Omega(r)$, where $r$ is
the distance to the central star. We make the assumption that the
disturbances due to thermal effects are small compared to the pressure
lengthscale $H$ of the disc. This assumption has been used and
discussed by \citet{2017MNRAS.472.4204M}. It allows us to perform our
analysis in a 3D shearing box \citep{shearingsheet}.
Our frame is therefore essentially a Cartesian box of dimensions much
smaller than the planet's semi-major axis, which contains the planet
and co-rotates with its guiding centre. We use the conventional
notation for the axes: $x$ is directed along the gradient of
unperturbed velocity, $y$ is directed along the unperturbed motion and
$z$ is perpendicular to the disc's midplane. We will refer to the
material at $x>0$ ($x<0$) as the outer (inner) disc, implying that the
central object lies on the negative side of the $x$-axis. We denote
with $E$ the epicyclic excursion of the planet: $E=ea$, where $e$ is
the eccentricity and $a$ the semi-major axis. Similarly, we denote
with $I$ its vertical excursion, and have $I=ia$, where $i$ is the
inclination. We note $\Omega_p$ the planet's orbital frequency and
restrict ourselves to the Keplerian case, for which the epicyclic
frequency matches the orbital frequency. In our frame, the planet
location is therefore:
\begin{equation}
  \label{eq:1}
  (x_p,y_p,z_p)=\left[x_p^0-E\cos(\Omega_pt), 2E\sin(\Omega_pt), I\sin(\Omega_p t')\right],
\end{equation}
where $x_p^0$ is the offset between the planet's guiding centre and its
corotation, $t$ is the time measured from a passage at periastron and
$t'=t-t_\mathrm{AN}$, where $t_\mathrm{AN}$ is a time of passage at
the ascending node.

Our governing equations are the continuity equation, Euler's
equations, and the equation on the internal energy density.
The continuity equation is:
\begin{equation}
  \label{eq:2}
\partial_t\rho + \bs{\nabla} .  \left( \rho \bs{V} \right) = 0
\end{equation}
where $\rho$ and $\bs{V} = (u,v,w)^T$ are respectively the density and the velocity of the gas. The Euler equation reads:
\begin{equation}
\label{eq:3}
\partial_t \bs V + \bs V.\bs{\nabla V} + 2\Omega_p \bs e_z \times \bs V = -\bs \nabla (\Phi_t+\Phi_p) - \frac{\bs \nabla p}{\rho},
\end{equation}
where $\Omega_p\bs e_z$ is rotation rate of the frame, $\Phi_p$ is the
planetary potential, $p$ the pressure and $\Phi_t$ is the tidal
potential given by:
\begin{equation}
  \label{eq:4}
  \Phi_t = -\frac 32\Omega_p^2(x-x_p^0)^2 + \frac{\Omega_p^2z^2}{2}.
\end{equation}
Finally, the equation for the density of internal energy, which we
denote with $\mathfrak e$ in order to avoid confusion with the
eccentricity, reads:
\begin{equation}
 \label{eq:5}
 \partial_t \mathfrak e + \bs \nabla.\left(\mathfrak e\bs V \right) = -p\bs \nabla . \bs V - \bs\nabla. \bs F_H+S, 
 \end{equation}
 where $S=S_d(\bs r) + S_p(\bs r)$ is a source term arising from the
 disc local heating $S_d(\bs r)$ and from the release of
 energy into the gas by the planet $S_p(\bs r)$. In what follows we
 assume the disc to be inviscid and 
 neglect the perturbation of $S_d(\bs r)$. We discuss in
 Appendix~\ref{sec:extens-visc-disc} how to extend our results to
 laminar, viscous discs. The heat flux
 $\bs F_H$ is given by:
\begin{equation}
  \label{eq:6}
  \bs F_H=-\rho\chi \bs\nabla\left(\frac{\mathfrak e}{\rho}\right),
\end{equation}
where  $\chi$ is the thermal diffusivity.

Hereafter we consider the perturbations of density $(\rho ')$,
velocities $(u', v', w')$ and pressure ($p'$) and we express the
perturbed quantities as the sum of the unperturbed quantities (with a
$0$ subscript) and the corresponding perturbations (primed) as:
\begin{eqnarray}
\rho &=& \rho_0 + \rho'\label{eq:7}\\
\mathfrak e &=& \mathfrak e_0 +\mathfrak e'\label{eq:8}\\
p &=& p_0 + p'\label{eq:9}\\
u &=& u'\label{eq:10}\\
v &=& v_0 + v' \equiv -\frac{3}{2}\Omega_p x + v' \label{eq:11}\\
w &=& w'\label{eq:12}.
\end{eqnarray}
The unperturbed velocity in Eq.~\eqref{eq:11} corresponds to the
Keplerian shear and comes from Eq.~\eqref{eq:3} for the unperturbed
quantities, which also yields:
\begin{equation}
\label{eq:13}
x_p^0 = -\frac{\partial_x p_0}{3\Omega_p^2\rho_0}.
\end{equation}
The offset between the guiding centre and corotation depends on the
pressure gradient. In what follows we assume $x_p^0\ll E$ and
$x_p^0\ll I$. We discuss how our results are affected when we relax
this hypothesis in appendix~\ref{sec:dist-from-guid}. We also assume
that the epicyclic and vertical excursions $E$ and $I$ are much
smaller than the characteristic size $\lambda$ of the disturbance:

\begin{equation}
  \label{eq:14}
  E \ll \lambda \mbox{~~~and~~~} I \ll \lambda.
\end{equation}

We consider the gas to be ideal. It obeys the relationship:
\begin{equation}
  \label{eq:15}
  p=(\gamma - 1)\mathfrak e, 
\end{equation}
where $\gamma$ is the adiabatic index. We denote with $c_s$ the
adiabatic sound speed, given by:
\begin{equation}
  \label{eq:16}
  c_s=\gamma p_0/\rho_0
\end{equation}
Upon linearization in $\rho'$, $u'$, $v'$, $w'$ and $\mathfrak e'$,
the governing equations can be recast as follows. The continuity
equation reads:
\begin{equation}
  \label{eq:17}
  \partial_t\rho'-\frac{3}{2}\Omega_px\partial_y\rho'+\rho_0(\partial_xu'+\partial_yv'+\partial_zw')=0,
\end{equation}
the three components of the Euler equation are:
\begin{equation}
  \label{eq:18}
  \partial_tu'-\frac{3}{2}\Omega_px\partial_yu'-2\Omega_pv'=-\partial_x\Phi_p-\frac{\partial_xp'}{\rho_0}+\frac{(\partial_xp_0)\rho'}{\rho_0^2},
\end{equation}
\begin{equation}
  \label{eq:19}
  \partial_tv'-\frac{3}{2}\Omega_px\partial_yv'+\frac{1}{2}\Omega_pu'=-\partial_y\Phi_p-\frac{\partial_yp'}{\rho_0},
\end{equation}
\begin{equation}
  \label{eq:20}
  \partial_tw'-\frac{3}{2}\Omega_px\partial_yw'=-\partial_z\Phi_p-\frac{\partial_zp'}{\rho_0}+\frac{(\partial_zp_0)\rho'}{\rho_0^2},
\end{equation}
and the energy equation becomes:
\begin{equation}
  \label{eq:21}
  \begin{split}
  \partial_tp' -\frac{3}{2}\Omega_px\partial_yp'+\gamma 
  p_0(\partial_xu'+\partial_yv'+\partial_zw')=\\\chi\Delta 
  p'-\chi\frac{p_0}{\rho_0}\Delta\rho'+(\gamma-1)S_p,
  \end{split}
\end{equation}
where $\Delta\equiv \partial^2_{x^2}+\partial^2_{y^2}+\partial^2_{z^2}$
is the Laplacian operator.
Following \citet{2017MNRAS.472.4204M}, we neglect the last term of
the right-hand side (R.H.S.) of Eq.~\eqref{eq:18} and  the third
term of the R.H.S. of Eq.~\eqref{eq:20}, as the size of the
perturbation is assumed small compared to the pressure lengthscale.

We perform a Fourier transform in the $y$ and $z$ directions with the
following convention in sign and normalization \footnote{Throughout 
  this work, we denote with $j$ the imaginary number with positive 
  imaginary part such that $j^2=-1$, so as to avoid confusion with the 
  inclination.}:
\begin{eqnarray}
  \tilde\xi(x,k_y,k_z) = \int\!\!\!\int \xi'(x,y,z)e^{-j(k_yy + k_zz)}dydz\, ,\label{eq:22}\\
  \xi'(x,y,z) = \frac{1}{4\pi^2}\int\!\!\!\int \tilde\xi(x,k_y,k_z)e^{j(k_yy + k_zz)}dk_ydk_z,\label{eq:23}
\end{eqnarray}
where $\xi'$ is the perturbation of any variable and $\tilde\xi$ its
Fourier transform. The derivative operators become:
\begin{eqnarray}
  \partial_x &\rightarrow &\partial_x\label{eq:24}\\
  \partial_y &\rightarrow& jk_y\label{eq:25}\\
  \partial_z &\rightarrow& jk_z\label{eq:26}\\
  \bs\nabla &\rightarrow& \tilde{\bs\nabla}  = (\partial_x, jk_y, jk_z)^T\label{eq:27}\\
  \Delta &\rightarrow& \Delta' = \partial_x^2 - k_y^2 -k_z^2 = \partial_x^2 - \bs k^2,\label{eq:28}
\end{eqnarray}
and Eqs.~(\ref{eq:17})-(\ref{eq:21}) can be rewritten as follows:
\begin{eqnarray}
\partial_t \tilde\rho-j\frac{3}{2}k_y\Omega_px\tilde\rho + \rho_0\tilde{\bs \nabla}.\tilde{\bs V} = 0  \label{eq:29}\\
\partial_t \tilde u-j\frac{3}{2}k_y\Omega_p x\tilde u - 2\Omega_p \tilde v + \frac{\partial_x \tilde p}{\rho_0} = -\partial_x \tilde\Phi_p\label{eq:30}\\
\partial_t \tilde v-j\frac{3}{2}k_y\Omega_p x\tilde v  + \frac{1}{2} \Omega_p\tilde u + \frac{jk_y \tilde p}{\rho_0} = -jk_y\tilde \Phi_p \label{eq:31}\\
\partial_t \tilde w-j\frac{3}{2}k_y\Omega_p x\tilde w +   \frac{jk_z \tilde p}{\rho_0} = -jk_z \tilde \Phi_p \label{eq:32} \\
\partial_t \tilde p-j\frac{3}{2}k_y\Omega_p x\tilde p + \gamma p_0 \tilde{\bs \nabla}.\tilde{\bs V} - \chi \Delta'\tilde p + \frac{\chi c_s^2}{\gamma}\Delta' \tilde\rho=(\gamma -1)\tilde S_p.  \label{eq:33}
\end{eqnarray}
Using Eq.~\eqref{eq:29} to substitute the velocity divergence
($\tilde{\bs \nabla}.\tilde{\bs V}$) in Eq.~\eqref{eq:33} we get:
\begin{equation}
\label{eq:34}
\left( \partial_t-j\frac{3}{2}k_y\Omega_p x \right)(\tilde p -
c_s^2\tilde\rho)  - \chi\Delta' \left(\tilde p 
  -\frac{c_s^2}{\gamma}\tilde\rho\right) = (\gamma -1)\tilde 
S_p. 
\end{equation}
This is our main equation. We work out a significant simplification in
the following section.

\section{Density response}
\label{sec:density-response}

%\subsection{Difference between the adiabatic case and the case with thermal diffusion for a cold planet}
%\label{sec:diff-betw-adiab}
\subsection{Simplification of the main equation}
\label{sec:simpl-main-equat}
In order to give an order of magnitude of the different terms that
feature in the density response, we take the Fourier transform in time
of the perturbation of the different quantities, and substitute the
partial derivative with respect to time in Eqs.~\eqref{eq:29}
to~\eqref{eq:32} by a multiplication by $-j\omega$.
Using Eqs.~\eqref{eq:30}-\eqref{eq:32}, we can write the three
components of the velocity in terms of $\tilde\Phi_p+\tilde p/\rho_0$.
Upon substitution in Eq.~\eqref{eq:29}, we are led to:
\begin{equation}
\label{eq:35}
\mathcal{K} (\tilde \rho)=  \mathcal{L}\left(\tilde \Phi_p + \frac{\tilde p}{\rho_0} \right), 
\end{equation}
where the linear operators ${\cal K}$ and ${\cal L}$ are defined
respectively by:
\begin{eqnarray}
\mathcal{K} (\tilde \rho) = \frac{\tilde \rho}{\rho_0} \!\!\! &+&\!\!\!   \left[  \frac{3 k_y \tilde\omega}{\Omega_p^3 D_2^2}\frac{\partial_x p_0}{\rho_0^2} - 
\frac{2k_y}{\Omega_p \tilde\omega D_2}\frac{\partial_x p_0}{\rho_0} \right] \frac{\tilde \rho}{\rho_0} - \nonumber\\
&&\frac{1}{\Omega_p^2 D_2}\frac{\partial_x p_0}{\rho_0^2} \partial_x \tilde \rho, \label{eq:36}
\end{eqnarray}
and 
\begin{equation}
  \mathcal{L}(Y) =  - \frac{\partial^2_x  
                                    Y}{\Omega_p^2 D_2} + \frac{3 k_y \tilde\omega}{\Omega_p^3 D_2^2}\partial_x Y  +
  \left[  \frac{k_y^2}{\Omega_p^2 D_2} \left( 1 - \frac{6}{D_2} \right)   + \frac{k_z^2}{\tilde \omega^2 }\right] Y, \label{eq:37}
\end{equation} 
where $D_2$ is the non-dimensional quantity:
\begin{equation}
\label{eq:38}
D_2 = \frac{9}{4}x^2k_y^2 - 1 +3xk_y\frac{\omega}{\Omega_p} +\left(\frac{\omega}{\Omega_p} \right)^2, 
\end{equation}
and $\tilde \omega$ is the frequency in the frame of matter: 
\begin{equation}
\label{eq:39}
\tilde \omega = \omega + \frac{3}{2}\Omega_pxk_y. 
\end{equation}
Using Eq.~\eqref{eq:13}, the linear operator
$\mathcal{K}(\tilde\rho)$ can be recast as:
\begin{equation}
 \mathcal{K} (\tilde \rho) = \frac{\tilde \rho}{\rho_0} -  \frac{6 k_y x_p^0}{D_2}\left[\frac{\Omega_p}{\tilde \omega}  -\frac{3 \tilde\omega}{2\Omega_p D_2}  
  \right] \frac{\tilde \rho}{\rho_0} + 
 \frac{3}{\rho_0 D_2}x_p^0\partial_x \tilde \rho. \label{eq:40}
\end{equation} 
We now specify to the case of a perturbation not triggered by a 
gravitational potential, such as the one arising from the release of 
heat by the planet. In that case: 
\begin{equation}
  \label{eq:41}
  {\cal K}(\tilde\rho)=\mathcal{L}\left(\frac{\tilde p}{\rho_0}\right). 
\end{equation}
We can work out the order of magnitudes of the different terms of this
identity using the following approximations:
$|x|\sim |k_y^{-1}|\sim |k_z^{-1}|\sim \lambda \ll H$,
$|\partial_x \tilde p|\sim |\tilde p|/H$,
$|\partial_{x^2}^2 \tilde p|\sim |\tilde p|/H^2$. We then consider
three cases:
\begin{enumerate}
\item Over most of the perturbation, we have
  $\tilde\omega/\Omega_p=O(1)$ and $D_2=O(1)$. In that case the
  dominant term of Eq.~\eqref{eq:37} is the last one that has the
  order of magnitude $|\tilde p|/[\rho_0(\Omega_p^2\lambda^2)]$, while
  the dominant term of Eq.~\eqref{eq:40} is the first one. We then
  have:
  \begin{equation}
    \label{eq:42}
    |\tilde p|\sim\lambda^2\Omega_p^2|\tilde\rho| \ll H^2\Omega_p^2|\tilde\rho|,
  \end{equation}
  hence $\tilde p$ is negligible compared to $\tilde\rho c_s^2$.
\item Whenever $\Omega_p/\tilde\omega \gg 1$, the dominant term in
  Eq.~\eqref{eq:37} is the last one and has order of magnitude $\tilde
  p/(\rho_0\lambda^2\tilde\omega^2)$, whereas the dominant term of
  Eq.~\eqref{eq:40} is
  $\mathrm{max}[1,(x_p^0/\lambda\times\Omega_p/\tilde\omega]\tilde\rho/\rho_0$,
  hence:
  \begin{equation}
    \label{eq:43}
    |\tilde p|\sim|\tilde\rho|\mathrm{max}(\lambda^2\tilde\omega^2,
    x_p^0\lambda\Omega_p\tilde\omega)\ll |\tilde\rho| H^2\Omega_p^2
  \end{equation}
\item Whenever $D_2\ll 1$, the dominant term in Eq.~\eqref{eq:37} is
  $\tilde p/(\lambda\Omega_pD_2)^2$ and the dominant term in
  Eq.~\eqref{eq:40} is $\mathrm{max}[1,x_p^0/(\lambda
  D_2^2)]\tilde\rho/\rho_0$, therefore:
  \begin{equation}
    \label{eq:44}
    |\tilde p|\sim |\tilde\rho|\mathrm{max}(\lambda^2\Omega_p^2D_2^2,
    \lambda x_p^0\Omega_p^2)\ll |\tilde\rho| H^2\Omega_p^2.
  \end{equation}
\end{enumerate}
The relation $|\tilde p|\ll c_s^2|\tilde\rho|$ is therefore verified
by any disturbance not triggered by a gravitational potential,
i.e. that verifies Eq.~\eqref{eq:41}, when the size of the disturbance
is much smaller than the pressure lengthscale.

\subsection{Net thermal effects}
%\label{sec:net-thermal-effects}
Hereafter we show that the thermal can be decomposed into two
contributions: one that arises from the inclusion of thermal
diffusion, even if the planet is non-luminous, and another one that
arises from the planet's luminosity itself. For this purpose we
perform the following decomposition of the perturbation:
\begin{eqnarray}
  \label{eq:45}
  \tilde \rho = \tilde \rho_a + \tilde \rho_t,\\
  \label{eq:46}
\tilde p = \tilde p_a + \tilde p_t,
\end{eqnarray}
where the $a$ subscript refers to the adiabatic solution, that of
Eqs.~\eqref{eq:29} to~\eqref{eq:33} with $\chi=0$ and $\tilde S_p=0$,
while the quantities without subscript refer to the general solution
of these equations (i.e. with $\chi\ne 0$ and $\tilde S_p\ne 0$). In
Eqs.~\eqref{eq:45} and~\eqref{eq:46}, the quantities with a $t$
subscript, which stands as \emph{thermal}, therefore appear as the
difference between the adiabatic solution and the solution with heat
release ($\tilde S_p\ne 0$) and thermal diffusion ($\chi\ne 0$).
Since the thermal diffusivity neither appears in the operator
$\mathcal{K}$ nor $\mathcal{L}$, we have:
\begin{equation}
  \label{eq:47}
  \mathcal{K}(\tilde\rho_a)=\mathcal{L}\left(\phi_p+\frac{\tilde p_a}{\rho_0}\right)
\end{equation}
and we can write, subtracting Eq.~\eqref{eq:47} from Eq.~\eqref{eq:35}
and using the linearity of these operators:
\begin{equation}
  \label{eq:48}
  \mathcal{K}(\tilde\rho_t)=\mathcal{L}\left(\frac{\tilde p_t}{\rho_0}\right).
\end{equation}
The size of the thermal disturbance being assumed small compared to
the pressure lengthscale, we then have, as shown above:
\begin{equation}
  \label{eq:49}
  |\tilde p_t|\ll
  c_s^2|\tilde\rho_t|.
\end{equation}
Noting that Eq.~\eqref{eq:33} reduces to
\begin{equation}
  \label{eq:50}
  \tilde p_a=c_s^2\tilde\rho_a,
\end{equation}
we recast Eq.~\eqref{eq:34} using the decomposition of
Eqs.~\eqref{eq:45} and~\eqref{eq:46}:
\begin{equation}
  \label{eq:51}
  \left( \partial_t-j\frac{3}{2}k_y\Omega_p x \right)  
  \tilde\rho_t  - \frac\chi\gamma  
    \Delta' \tilde\rho_t\approx -\frac{\gamma -1}{c_s^2}\tilde 
  S_p-\chi\frac{\gamma-1}{\gamma c_s^2}\Delta'\tilde p_a,  
\end{equation} 
where we have used Eq.~\eqref{eq:49} to get rid of the instances of
$\tilde p_t$. The second term of the R.H.S. appears as an additional
source term. In order to proceed, we make the same approximation as
\citet{2017MNRAS.472.4204M} and write:
\begin{equation}
  \label{eq:52}
  \rho_0\tilde\Phi_p+\tilde p_a=0.
\end{equation}
This approximation is valid for low-mass planets
($M/M_\star\ll (H/r)^3$) which have a low, largely subsonic velocity
with respect to the ambient gas.  Assuming that the relative
variations of $\rho_0$ are much smaller than those of $\Phi_p$ over
the perturbed region, we have, going back to real space:
\begin{equation}
  \label{eq:53}
  \left( \partial_t-\frac{3}{2}\Omega_p x\partial_y \right) 
  \rho_t'  - \frac\chi\gamma  
    \Delta \rho_t'\approx -\frac{\gamma -1}{c_s^2}
  S_p+\chi\frac{(\gamma-1)\rho_0}{\gamma c_s^2}\Delta\Phi_p. 
\end{equation}
Using Poisson's equation for the planetary potential:
\begin{equation}
  \label{eq:54}
  \Delta\Phi_p=4\pi GM\delta(x-x_p)\delta(y-y_p)\delta(z-z_p),
\end{equation}
and specifying from now on to a singular planetary heating term:
\begin{equation}
  \label{eq:55}
  S_p=L\delta(x-x_p) \delta(y-y_p)\delta(z-z_p),
\end{equation}
where $L$ is the planet's luminosity, we can recast Eq.~\eqref{eq:53}
as
\begin{equation}
  \label{eq:56}
  \left( \partial_t-\frac{3}{2}\Omega_p x \partial_y\right) 
  \rho_t'  - \frac\chi\gamma  
    \Delta \rho_t'\approx -\frac{\gamma -1}{c_s^2}
  S^\mathrm{tot}_p, 
\end{equation}
where the modified source term $S^\mathrm{tot}_p$ is given by:
\begin{equation}
  \label{eq:57}
  S^\mathrm{tot}_p=(L-L_c)\delta(x-x_p)\delta(y-y_p)\delta(z-z_p),
\end{equation}
where $L_c$ has the value:
\begin{equation}
  \label{eq:58}
  L_c=\frac{4\pi GM\chi\rho_0}{\gamma}.
\end{equation}
In what follows, Eq.~\eqref{eq:56} or its Fourier transform in $y$ and
$z$, which reads:
\begin{equation}
  \label{eq:59}
  \left( \partial_t-\frac{3}{2}j\Omega_p xk_y \right) 
  \tilde\rho_t  - \frac\chi\gamma  
    \Delta'\tilde\rho_t\approx -\frac{\gamma -1}{c_s^2}
  \tilde S^\mathrm{tot}_p,
\end{equation}
constitutes our main equation.

\subsection{Evaluation of the source terms}
\label{sec:source-terms}
The source term is given by:
\begin{eqnarray}
 S^\mathrm{tot}_p(x,y,z)&=&(L-L_c)\delta(x-x_p)\delta(y-y_p)\delta(z-z_p) \nonumber\\
 &=&(L-L_c)\delta\left[x+(E\cos\Omega_p t)-x_p^0\right]\delta\left(y-2E\sin\Omega_p t\right) \nonumber\\
&&\times \delta\left(z - I\sin\Omega_p t'\right).\label{eq:60}
\end{eqnarray}
As said in section~\ref{sec:governing-equations}, we neglect from now
on $x_p^0$ with respect to $E$ and $I$. Taking the Fourier transform
in $y$ and $z$ of Eq.~\eqref{eq:60}, we arrive at:
\begin{eqnarray}
\tilde S^\mathrm{tot}_p(x,k_y,k_z) \!\!\!\! &=&\!\!\!\!  (L-L_c)\left[\delta(x)+E\cos\Omega_p t\delta'(x)\right] \times \nonumber\\
&&\!\!\!\!\left(1 -2jEk_y \sin\Omega_p t\right)  \left(1- jIk_z\sin\Omega_p t'\right)\label{eq:61}
\end{eqnarray}
which reads, keeping only the terms up to first order in $E$ and $I$
as per our hypothesis of Eq.~\eqref{eq:14}:
\begin{eqnarray}
\tilde S^\mathrm{tot}_p(x,k_y,k_z) &=&(L-L_c)\delta(x) +  (L-L_c)E\delta'(x)\cos\Omega_p t\nonumber\\
         &&-2jE(L-L_c)k_y \delta(x)\sin\Omega_p t\nonumber\\
         &&-jI(L-L_c)k_z\delta(x)\sin\Omega_p t'.\label{eq:62}
\end{eqnarray}
We recognize the source term of the circular, coplanar case ($E=I=0$)
in the first term of the R.H.S. The two subsequent terms arise from
the eccentricity and the last one from inclination. If we define:
\begin{equation}
\label{eq:63}
  \tilde S_e(x,k_y,k_z)=(L-L_c)\delta'(x)\cos\Omega_p t-2j(L-L_c)k_y\delta(x)\sin\Omega_pt
\end{equation}
and
\begin{equation}
  \label{eq:64}
 \tilde S_i(x,k_y,k_z)=-j(L-L_c)k_z\delta(x) \sin\Omega_p t',
\end{equation}
then we have:
\begin{equation}
  \label{eq:65}
  \tilde S^\mathrm{tot}_p(x,k_y,k_z)=\tilde S_c(x,k_y,k_z)+E\tilde S_e(x,k_y,k_z)
  +I\tilde S_i(x,k_y,k_z),
\end{equation}
where the ``circular'' term $S_c(x,k_y,k_z)$ is $(L-L_c)\delta (x)$. Since Eq.~\eqref{eq:59} is linear, we
have:
\begin{equation}
  \label{eq:66}
  \tilde\rho=\tilde\rho_c+E\tilde\rho_e+I\tilde\rho_i,
\end{equation}
where $\tilde\rho_c$ ($\tilde\rho_e$, $\tilde\rho_i$) is the
solution of Eq.~\eqref{eq:59} where the source term
$\tilde S^\mathrm{tot}_p$ of the R.H.S. is replaced by $\tilde S_c$
($\tilde S_e$, $\tilde S_i$).  The term $\tilde\rho_c$ is the response
to $\tilde S_c$ and has been studied by \citet{2017MNRAS.472.4204M}.
We focus in the following on the response to the ``eccentric'' and
``inclination'' source terms $\tilde S_e$ and $\tilde S_i$. We note
that in real space Eq.~\eqref{eq:66} is simply:
\begin{equation}
  \label{eq:67}
  \rho'=\rho_c'+E\rho_e'+I\rho_i',
\end{equation}
and we mention that $\rho_e'$ and $\rho_i'$ are not dimensionally
homogeneous to densities, as the factors $E$ and $I$ have unit of
length.

These source terms can be expressed in a slightly different manner. If
we define:
\begin{equation}
  \label{eq:68}
  S_e^\pm=(L-L_c)\left[\mp k_y\delta(x)+\frac{\delta'(x)}{2}\right]
\end{equation}
and
\begin{equation}
  \label{eq:69}
  S_i^\pm(x)=\pm\frac{(L-L_c)k_z}{2}\delta(x),
\end{equation}
then the source terms can be written respectively as:
\begin{equation}
  \label{eq:70}
  \tilde S_e=S_e^+\exp(j\Omega_pt)+S_e^-\exp(-j\Omega_pt)
\end{equation}
and
\begin{equation}
  \label{eq:71}
  \tilde S_i=S_i^+\exp(j\Omega_pt')+S_i^-\exp(-j\Omega_pt').
\end{equation}
We denote with $\tilde\rho_e^\pm\exp(\pm j\Omega_pt)$ the solution to
Eq.~\eqref{eq:59} where the source term of the R.H.S. is replaced by
$S_e^\pm\exp(\pm j\Omega_pt)$. Similarly, we denote with
$\tilde\rho_i^\pm\exp(\pm j\Omega_pt')$ the solution of the same
equation where the source term of the R.H.S is replaced by
$S_i^\pm\exp(\pm j\Omega_p t')$. We therefore have:
\begin{equation}
  \label{eq:72}
  j\Omega_p\left(1-\frac32 xk_y\right)\tilde\rho_e^+=\frac{\chi}{\gamma}\Delta'\tilde\rho_e^+-\frac{(\gamma-1)S_e^+}{c_s^2}
\end{equation}
and
\begin{equation}
  \label{eq:73}
  j\Omega_p\left(-1-\frac32 xk_y\right)\tilde\rho_e-=\frac{\chi}{\gamma}\Delta'\tilde\rho_e^-+\frac{(\gamma-1)S_e^-}{c_s^2}
\end{equation}
Noting that $S_e^+(x)=-S_e^-(-x)$, it is straightforward to show that:
\begin{equation}
  \label{eq:74}
  \tilde\rho_e^+(x)=-[\tilde\rho_e^-(-x)]^*,  
\end{equation} 
where the $*$ superscript denotes the complex conjugate. We therefore
focus only on the solution of Eq.~\eqref{eq:72} and will use
Eq.~\eqref{eq:74} whenever the expression of $\tilde\rho_e^-$ is
needed. In a similar manner, we have:
\begin{equation}
  \label{eq:75}
  j\Omega_p\left(1-\frac32 xk_y\right)\tilde\rho_i^+=\frac{\chi}{\gamma}\Delta'\tilde\rho_i^+-\frac{(\gamma-1)S_i^+}{c_s^2},
\end{equation}
and we use the relationship
\begin{equation}
  \label{eq:76}
  \tilde\rho_i^+(x)=-[\tilde\rho_i^-(-x)]^*, 
\end{equation}
to infer $\tilde\rho_i^-$ whenever needed.

\subsection{Non-dimensional form of the main equation}
\label{sec:non-dimensional-form}
We write Eqs.~\eqref{eq:72} and~\eqref{eq:75} in non-dimensional
form. For that purpose we introduce new variables:
\begin{equation}
  \label{eq:77}
  x_c=\frac{2}{3 k_y},
\end{equation}
\begin{equation}
  \label{eq:78}
  K=\frac{2 \chi k^3}{3\gamma \Omega_p k_y},
\end{equation}
\begin{equation}
  \label{eq:79}
  X=xk,\mbox{~~~~and~~~~}X_c=x_ck,
\end{equation}
where $k=(k_y^2+k_z^2)^{1/2}$. This allows us to rewrite
Eq.~\eqref{eq:72} as:
\begin{equation}
  \label{eq:80}
 j(X_c-X)\tilde\rho_e^+=K(\tilde\rho_e^{+ ''}-\tilde\rho_e^+) + s_1\left[\delta(X)-\frac34X_c\delta'(X)\right], 
\end{equation}
where $\tilde\rho_e^{+ ''}$ is the second derivative of
$\tilde\rho_e^+$ with respect to $X$ and where:
\begin{equation}
  \label{eq:81}
  s_1=\frac{2(\gamma-1)k^2(L-L_c)}{3\Omega_p c_s^2}.
\end{equation}
Similarly, we have:
\begin{equation}
\label{eq:82}
  j(X_c-X)\tilde\rho_i^+=K(\tilde\rho_i^{+ ''}-\tilde\rho_i^+) + s_2\delta(X),
\end{equation}
where:
\begin{equation}
  \label{eq:83}
 s_2=\frac{(\gamma-1)k^2k_z(L-L_c)}{3\Omega_p k_yc_s^2}.
\end{equation}

We call $T_{X_c,K}$ ($W_{X_c,K}$) the solution of Eq.~\eqref{eq:80}
[Eq.~\eqref{eq:82}] with $s_1=1$ ($s_2=1$)
whose real and imaginary parts tend to zero in $\pm\infty$, and we
have:
\begin{equation}
  \label{eq:84}
  \tilde\rho_e^+=s_1T_{X_c,K}
\end{equation}
and
\begin{equation}
  \label{eq:85}
  \tilde\rho_i^+=s_2W_{X_c,K}
\end{equation}
We describe hereafter how we obtain the solutions $T_{X_c,K}$ and $W_{X_c,K}$.

\subsection{Jump conditions at origin}
\label{sec:jump-conditions}
From the considerations above, the function $T(X;X_c,K):\mathbb{R}\rightarrow \mathbb{C}$ verifies: 
\begin{equation}
  \label{eq:86}
  j(X_c-X)T=K(T''-T)+\delta(X)-3X_c\delta'(X)/4,  
\end{equation} 
where for the sake of clarity we omit the full dependence of $T$.  
We work out the jump conditions that $T$ must fulfil  
in $X=0$. We assume $T$ is finite (but possibly 
discontinuous). We call $F_\varepsilon(X)$ the following integral of 
$j(X_c-X)T+KT$, where $\varepsilon>0$: 
\begin{equation}
  \label{eq:87}
  F_\varepsilon(X)=\int_{-\varepsilon}^Xj(X-X_c)T+KTdX. 
\end{equation}
Since the integrand is finite, $F$ is a continuous function of $X$ or $\varepsilon$ in 
zero. We have, using Eq.~\eqref{eq:86}: 
\begin{equation}
  \label{eq:88}
  F_\varepsilon(X)=K[T'(X)-T'(-\varepsilon)]+H(X)-3X_c\delta(X)/4, 
\end{equation}
where $H(X)$ is Heaviside's step function.
We now define: 
\begin{equation}
  \label{eq:89}
  G_\varepsilon(X)=\int_{-\varepsilon}^XF_\varepsilon(X)+KT'(\varepsilon)-H(X)\,dX, 
\end{equation}
which, like $F$, is a continuous function of $X$ or $\varepsilon$ in 
zero. We have: 
\begin{equation}
  \label{eq:90}
  G_\varepsilon(X)=K[T(X)-T(-\varepsilon)]-3X_cH(X)/4. 
\end{equation}
Evaluating Eq.~\eqref{eq:88} in $X=\varepsilon$ and letting 
$\varepsilon\rightarrow 0$, we get: 
\begin{equation}
  \label{eq:91}
  T'(0^+)-T'(0^-)=-\frac 1K, 
\end{equation} 
while evaluating Eq.~\eqref{eq:90} in $X=\varepsilon$ and letting 
$\varepsilon\rightarrow 0$, we get: 
\begin{equation}
  \label{eq:92}
  T(0^+)-T(0^-)=\frac{3X_c}{4K}. 
\end{equation}
Eqs.~\eqref{eq:91} and~\eqref{eq:92} are the jump conditions in
$X=0$. As their R.H.S. are real, the imaginary part of the solution
and of its derivative are continuous in $X=0$, whereas their real
parts undergo the jumps above.  When $X_c=0$, Eq.~\eqref{eq:86}
reduces to:
\begin{equation}
  \label{eq:93}
  -jXT=K(T''-T)+\delta(X), 
\end{equation}
which is the equation of the circular problem, and has solution $R_K+jI_K$ \citep[see][Eqs.~(75)--(79)]{2017MNRAS.472.4204M}.
Similarly, the function $W(X;X_c,K):\mathbb{R}\rightarrow \mathbb{C}$
verifies the differential equation: 
\begin{equation}
  \label{eq:94}
  j(X_c-X)W=K(W''-W)+\delta(X),
\end{equation}
and can be easily shown to be continuous in $X=0$, while its first
derivative has the following jump:
\begin{equation}
  \label{eq:95}
  W'(0^+)-W'(0^-)=-\frac 1K.
\end{equation}

\subsection{Numerical procedure}
\label{sec:resolution-practice}
We use a shooting method with a Runge-Kutta scheme at fifth order and
start our integration at a large distance $\pm X_0$ from the
planet. We integrate towards corotation ($X=0$), so that the
integration started in $-X_0$ goes forward and that started in $X_0$
goes backwards. Our boundary conditions in $\pm X_0$ consist of the
real and imaginary parts of the solution. Its first-order derivative
can be chosen as in \citet{2017MNRAS.472.4204M} (Eqs.~A3 and~A4), but
we find that our solution is insensitive to this choice (provided
$X_0$ is large enough), and simpler choices, such as a null
derivative, yield virtually the same solution. In the following we
present the method employed to obtain the solution $T_{X_c,K}$. It can
be applied straightforwardly to $W_{X_c,K}$ by amending the jump
conditions in $X=0$. We denote with $\bs I= \{\Re [T(X_0)], \Im [T(X_0)],
\Re [T(-X_0)], \Im [T(-X_0)]\}^T$ our input vector. Upon integration, we
obtain the jumps in $X=0$ corresponding to our choice of $\bs
I$. These jumps constitute our output vector:
\begin{equation}
  \label{eq:96}
  \bs O= \{\Delta\Re(T), \Delta\Im(T),
\Delta\Re(T'), \Delta\Im(T')\}^T,
\end{equation}
where, here only, the $\Delta$ symbol represents the difference in $X=0$
between the values numerically obtained by the backwards integration
with $X\ge 0$ and the forward integration with $X\le 0$. Our aim is
that the output vector matches the jump conditions, i.e.
\begin{equation}
  \label{eq:97}
  \bs O = \bs J,
\end{equation}
with:
\begin{equation}
  \label{eq:98}
  \bs J = (3X_c/4K, 0, -1/K, 0)^T,
\end{equation}
as required by Eqs.~\eqref{eq:91} and~\eqref{eq:92} and the fact that
the imaginary part of the solution and its first derivative is
continuous in $X=0$. Since Eq.~\eqref{eq:86} is linear in $T$ over the
domains $X>0$ and $X<0$, the output vector is a linear function of the
input vector, i.e. there exists a $4\times 4$ matrix ${\cal M}$ with
real coefficients such that:
\begin{equation}
  \label{eq:99}
  \bs O = {\cal M}\bs I.
\end{equation}
We construct the matrix ${\cal M}$ column by column, using successively
$\bs I=(1,0,0,0)^T$, $\bs I=(0,1,0,0)^T$, $\bs I=(0,0,1,0)^T$ and
$\bs I=(0,0,0,1)^T$. The first two cases correspond to $T(-X_0)=0$
and $T(X_0)=1$ or $T(X_0)=j$, respectively, while the last two cases
correspond to $T(X_0)=0$ and $T(-X_0)=1$ or $T(-X_0)=j$, respectively.
The output vectors give directly the columns of the matrix ${\cal
  M}$. Once the latter is built, we determine the boundary conditions
$\bs I_T$ for the sought solution $T$ using:
\begin{equation}
  \label{eq:100}
  \bs I_T={\cal M}^{-1}\bs J.
\end{equation}
The solution $T$ is then constructed by a fifth and last integration
using the boundary conditions $\bs I_T$. It fulfils the jump
conditions in $X=0$ by construction. The solution
$W_{X_c,K}$ must verify the following differential equation:
\begin{equation}
  \label{eq:101}
    j(X_c-X)W=K(W''-W)+\delta(X).
\end{equation}
It is constructed in almost the same manner, except that we use
different jump conditions given by:
\begin{equation}
  \label{eq:102}
  \bs J'=(0, 0, -1/K, 0)^T.
\end{equation}

%%%%%%%%%%%%%%%%%%%%%%%%%%%
%%%%%%%%%%%%%%%%%%%%%%%%%%%
%%%%           Force expression.        %%%%%%%%
%%%%%%%%%%%%%%%%%%%%%%%%%%%
%%%%%%%%%%%%%%%%%%%%%%%%%%%

\section{Force expression}
\label{sec:force-expression}
Having determined the density response, we can now calculate the force
exerted on the planet by the perturbed density. Its three components read:
 \begin{eqnarray}
F_x&=&\int\!\!\int\!\!\int \rho'\partial_x\Phi_p\,dx \,dy\,dz,\label{eq:103}\\
F_y&=&\int\!\!\int\!\!\int \rho'\partial_y\Phi_p\,dx \,dy\,dz, \label{eq:104}\\
F_z&=&\int\!\!\int\!\!\int \rho'\partial_z\Phi_p\,dx \,dy\,dz. \label{eq:105}
\end{eqnarray}
The gravitational potential of the planet is:
\begin{eqnarray}
\label{eq:106}
\Phi_p(x, y, z) =- G M\times \left[(x+E\cos\Omega_p
      t)^2 +\right. \nonumber \\
      \left. (y-2E\sin\Omega_p t)^2+(z - I\sin\Omega_p t')^2\right]^{-1/2},
\end{eqnarray}
which reads at first order in $E$ and $I$:
\begin{equation}
  \label{eq:107}
  \Phi_p(x,y,z)=\Phi_c(x,y,z)+E\Phi_e(x,y,z)+I\Phi_i(x,y,z),
\end{equation}
with
\begin{eqnarray}
\label{eq:108}
  \Phi_c(x,y,z) &=& -\frac{G M}{R}\\
  \label{eq:109}
  \Phi_e(x,y,z) &=& \frac{G M}{R^3} (x\cos\Omega_pt- 2y\sin\Omega_pt)\\
  \label{eq:110}
  \Phi_i(x,y,z) &=& -\frac{G M}{R^3} (z\sin\Omega_p t'),
\end{eqnarray}
where $R=(x^2+y^2+z^2)^{1/2}$. In what follows we first work out a
separation of the terms into those arising from the eccentricity,
which give rise to the horizontal components of the force, and those
arising from the inclination, which give rise to the vertical
component of the force, then we carry out the calculation of the force
components.

\subsection{Separation of eccentricity and inclination contributions}
\label{sec:separ-eccentr-incl}
Substituting Eqs.~\eqref{eq:67} and~\eqref{eq:107} in
Eq.~\eqref{eq:103}, we are led to:
\begin{equation}
  \label{eq:111}
  F_x=\iiint(\rho'_c+E\rho'_e+I\rho'_i)\partial_x(\Phi_c+E\Phi_e+I\Phi_i) dxdydz.
\end{equation}
Expanding this expression to first order in $E$ and $I$, we are left
with:
\begin{eqnarray}
  F_x&=\iiint&\left[\rho'_c\partial_x\Phi_c +
  E(\rho'_e\partial_x\Phi_c+\rho'_c\partial_x\Phi_e)\right.\nonumber\\
  &&\left.+I(\rho'_i\partial_x\Phi_c+\rho'_c\partial_x\Phi_i)\right]dxdydz.   \label{eq:112}
\end{eqnarray}
Since Eq.~\eqref{eq:59} does not include partial derivatives in $z$ of
odd order, the solution has the same parity in $z$ as the source
term. Since $S_e$ is even in $z$, so is $\rho'_e$, and since $S_i$ is
odd in $z$, so is $\rho'_i$. Furthermore, Eqs.~\eqref{eq:109}
and~\eqref{eq:110} show respectively that $\Phi_e$ is even in $z$ and
$\Phi_i$ is odd in $z$. This implies that the part of the integrand in
factor of $I$ in Eq.~\eqref{eq:112} vanishes: the force in the $x$
direction depends only on the eccentricity. The same is true of the
force in the $y$-direction. On the other hand, since the derivative in
$z$ changes the parity in $z$ of the function to which it is applied,
the component $F_z$ of the force reduces to:
\begin{equation}
  \label{eq:113}
  F_z=\iiint I(\rho'_i\partial_z\Phi_c+\rho'_c\partial_z\Phi_i)\,dx\,dy\,dz.
\end{equation}
The vertical component of the force therefore only depends on the
inclination. In this section we have used the expansion of the
planetary potential in real space to separate the force expression
into contributions arising from the eccentricity or the
inclination. In what follows, we evaluate the force using the
expression of the potential in Fourier space, using for that purpose
Parseval-Plancherel's theorem, which reads, with the conventions of
Eqs.~\eqref{eq:22} and~\eqref{eq:23}:
\begin{equation}
  \label{eq:114}
  \iint f'(y,z)g'(y,z)dy\,dz=\frac{1}{4\pi^2}\iint \tilde
  f(k_y,k_z)\tilde g^*(k_y,k_z)dk_y\,dk_z.
\end{equation}

\subsection{Force component in the $x$-direction}
\label{sec:radial-force}
Using Eqs.~\eqref{eq:112} and~\eqref{eq:114} we can write the
$x$-component of the force as:
\begin{equation}
  \label{eq:115}
  F_x=F_x^{(1)}+F_x^{(2)}
\end{equation}
with
\begin{equation}
  \label{eq:116}
  F_x^{(1)}=\frac{E}{4\pi^2}\iiint\tilde\rho_e\partial_x\tilde\Phi_c^*dx\,dky\,dkz,  
\end{equation}
and
\begin{equation}
  \label{eq:117}
  F_x^{(2)}=\frac{E}{4\pi^2}\iiint\tilde\rho_c\partial_x\tilde\Phi_e^*dx\,dky\,dkz. 
\end{equation} 
We have omitted the constant term $\rho_c\partial_x\Phi_c$,
which vanishes in the shearing sheet for symmetry reasons. The
expression of $\tilde\Phi_c$ is \citep[][Eq.~40]{2017MNRAS.472.4204M}:
\begin{equation}
 \label{eq:118}
  \tilde\Phi_c(x,k_y,k_z)=-\frac{2\pi GM}{k}\exp(-k|x|).
\end{equation}
The expression of the potential in the eccentric case can be inferred
from Eq.~\eqref{eq:118} through a shift in $x$ and $y$:
\begin{equation}
  \label{eq:119}
  \begin{split}
  \tilde\Phi(x,k_y,k_z)=-\frac{2\pi
    GM}{k}\exp(-k|x+E\cos\Omega_pt|)\\
  \times\exp(-2jEk_y\sin\Omega_pt).
  \end{split}
\end{equation}
Using our hypothesis of Eq.~\eqref{eq:14}, we expand this expression to first order in $E$. This yields:
\begin{equation}
  \label{eq:120}
  \tilde\Phi=\tilde\Phi_c+E\tilde\Phi_e^+\exp(j\Omega_pt)+E\tilde\Phi_e^-\exp(-j\Omega_pt),
\end{equation}
where the dependence on $x$, $k_y$ and $k_z$ has been omitted for
improved legibility, and where:
\begin{equation}
\label{eq:121}
  \tilde\Phi_e^+(x,k_y,k_z)=-\frac{2\pi 
    GM}{k}\exp(-k|x|)\left[- k_y-\frac k2\text{sgn}(x)\right],
\end{equation}
and
\begin{equation}
\label{eq:122}
  \tilde\Phi_e^-(x,k_y,k_z)=-\frac{2\pi 
    GM}{k}\exp(-k|x|)\left[k_y-\frac k2\text{sgn}(x)\right].
\end{equation}
Using Eqs.~\eqref{eq:116} and~\eqref{eq:118}, we have:
\begin{equation}
  \label{eq:123}
  \begin{aligned}
  F_x^{(1)}=\frac{GME}{2\pi}\iiint  & [\tilde\rho_e^+ \exp(j\Omega_p t)+\tilde\rho_e^-\exp(-j\Omega_p
  t)]\times\\
  & \exp(-k|x|)\text{sgn}(x) dx d^2\bs k
  \end{aligned}
\end{equation}
Using Eq.~\eqref{eq:74}, and performing the change of variable
$x\rightarrow -x$ on one of the terms of the integrand, this
expression can be recast as:
\begin{equation}
  \label{eq:124}
  F_x^{(1)}=\frac{GME}{\pi}\iiint\Re[\tilde\rho_e^+\exp(j\Omega_p
  t)]\exp(-k|x|)\text{sgn}(x)dx d^2\bs k.
\end{equation}
We start the evaluation of $F_x^{(2)}$ using an integration by parts,
which leads, using Eq.~\eqref{eq:117}:
\begin{equation}
  \label{eq:125}
  F_x^{(2)}=-\frac{E}{4\pi^2}\iiint(\partial_x\tilde\rho_c)\tilde\Phi_e^*dx\,d^2\bs k. 
\end{equation}
We note that $\tilde\Phi_e^+(-x)=-\tilde\Phi_e^-(x)$ and that
\citep[][Eq.~116 and Appendix~A]{2017MNRAS.472.4204M}:
\begin{equation}
  \label{eq:126}
  \partial_x\tilde\rho_c(-x)=-[\partial_x\tilde\rho_c(x)]^*,
\end{equation}
so that we can use a change of variable similar to that used in
evaluating $F_x^{(1)}$. This yields:
\begin{equation}
  \label{eq:127}
  F_x^{(2)}=-\frac{E}{2\pi^2}\iiint\Re\left[\partial_x\tilde\rho_c\exp(-j\Omega_pt)\right]\Phi_e^+\,dx\,d^2\bs k.
\end{equation}
Using Eqs.~\eqref{eq:115}, \eqref{eq:121}, \eqref{eq:124}
and~\eqref{eq:127}, we can write:
\begin{equation}
  \label{eq:128}
  F_x=F_x^C\cos\Omega_p t+F_x^S\sin\Omega_p t
\end{equation}
with:
\begin{equation}
  \label{eq:129}
  F_x^C=\frac{GME}{\pi}\iiint\left(\tilde\rho_e^R-\frac{\partial_x\tilde\rho_c^R}{2}\right)\exp(-k|x|)
  \text{sgn}(x) dx\,d^2\bs k
\end{equation}
and:
\begin{equation}
  \label{eq:130}
  F_x^S=\frac{GME}{\pi}\iiint\left(-\tilde\rho_e^I-
    k_y\tilde\rho_c^I\right)\exp(-k|x|)\text{sgn}(x)dx\,d^2\bs k,
\end{equation}
where for a more concise notation we define
$\tilde\rho_e^R\equiv\Re(\tilde\rho_e^+)$,
$\tilde\rho_c^R\equiv\Re(\tilde\rho_c)$,
$\tilde\rho_e^I\equiv\Im(\tilde\rho_e^+)$ and
$\tilde\rho_c^I\equiv\Im(\tilde\rho_c)$.  An integration by parts has been
used to write the second term of the integrand of Eq.~\eqref{eq:130}.

\subsection{Force component in the $y$-direction}
\label{sec:azim-comp-force}
The $y$-component of the force is calculated following an approach
very similar to that of section~\ref{sec:radial-force}. It is
expressed as:
\begin{equation}
  \label{eq:131}
  F_y=F_y^{(1)}+F_y^{(2)}
\end{equation}
with
\begin{equation}
  \label{eq:132}
  F_y^{(1)}=-\frac{E}{4\pi^2}\iiint jk_y \left(\tilde\rho_e^+e^{j\Omega_p 
  t}
  +\tilde\rho_e^-e^{-j\Omega_p t}\right)\tilde\Phi_c dx\,d^2\bs k  
\end{equation} 
and  
\begin{equation}
  \label{eq:133}
  F_y^{(2)}=-\frac{E}{4\pi^2}\iiint\!  
  jk_y\tilde\rho_c\left(\tilde\Phi_e^+e^{-j\Omega_pt}
    +\tilde\Phi_e^-e^{j\Omega_pt}\right)dx\,d^2\bs  
  k.  
\end{equation} 
Eq.~\eqref{eq:132} is transformed in a similar way as the expressions
of $F_x^{(1)}$ and $F_x^{(2)}$. We use the fact that $\tilde\Phi_c$ is
even in $x$, and make use again of Eq.~\eqref{eq:74} to write:
\begin{equation}
  \label{eq:134}
  F_y^{(1)}=\frac{E}{2\pi^2}\iiint\Im[\tilde\rho_e^+\exp(j\Omega_p 
  t)]k_y\tilde\Phi_cdx\,d^2\bs k. 
\end{equation}
Using Eqs.~\eqref{eq:121} and~\eqref{eq:122}, we can rewrite
Eq.~\eqref{eq:133} as:
\begin{equation}
  \label{eq:135}
  \begin{split}
  F_y^{(2)}=-\frac{GME}{\pi}&\iiint \left[\frac{k_y}{k}\sin(\Omega_pt)+\right.\\
&\left.    \frac{j}{2} 
    \cos(\Omega_pt)\mathrm{sgn}(x)\right]e^{-k|x|} k_y\tilde\rho_cdx\,d^2\bs k. 
\end{split}
\end{equation}
Since $\tilde\rho_c^R$ ($\tilde\rho_c^I$) is even (odd) in $x$, the
contribution of the imaginary part of $\tilde\rho_c$ to the first term
of the integral vanishes, and so does the contribution of its real
part to the second term. From Eqs.~\eqref{eq:134} and~\eqref{eq:135}
we can write:
\begin{equation}
  \label{eq:136}
  F_y=F_y^C\cos\Omega_p t+F_y^S\sin\Omega_p t,  
\end{equation}
where the two components read respectively:
\begin{equation}
  \label{eq:137}
F_y^C=\frac{E}{2\pi^2}\iiint
k_y\tilde\Phi_c\left(\tilde\rho_e^{I,+}-\partial_x\tilde\rho_c^I/2\right)dx\,d^2\bs k
\end{equation}
and:
\begin{equation}
  \label{eq:138}
  F_y^S=\frac{E}{2\pi ^2}\iiint 
  k_y\tilde\Phi_c\left(\tilde\rho_e^{+,R}+k_y\tilde\rho_c^R\right)dx\,d^2\bs k,
\end{equation}
where we have performed an integration by parts to write the second
term of the integrand of $F_y^C$.

\subsection{Vertical component of the force}
\label{sec:vert-comp-force}
Prior to the evaluation of the force, we need to work out the Fourier
transform of the potential to first order in $I$, using our hypothesis
of Eq.~\eqref{eq:14}. It can be obtained from Eq.~\eqref{eq:118}
through a shift of magnitude $I\sin\Omega_p t'$ in the
$z$~direction. It reads:
\begin{equation}
  \label{eq:139}
  \tilde\Phi=\tilde\Phi_c+I\tilde\Phi_i^+\exp(j\Omega t')+I\tilde\Phi_i^-\exp(-j\Omega t')  
\end{equation} 
where  
\begin{equation}
\label{eq:140}
 \tilde\Phi_i^\pm(x,k_y,k_z) = \pm\frac{2\pi G M}{k}\exp(-k|x|)\frac{k_z}{2},  
\end{equation}
As done previously with the other components, we split the
$z$-component of the force into two contributions:
\begin{equation}
  \label{eq:141}
  F_z=F_z^{(1)}+F_z^{(2)},
\end{equation}
with
\begin{equation}
  \label{eq:142}
  F_z^{(1)}=-\frac{I}{4\pi^2}\iiint jk_z \left(\tilde\rho_i^+e^{j\Omega_p 
  t'}
  +\tilde\rho_i^-e^{-j\Omega_p t'}\right)\tilde\Phi_c dx\,d^2\bs k  
\end{equation} 
and  
\begin{equation}
  \label{eq:143}
  F_z^{(2)}=-\frac{I}{4\pi^2}\iiint\!  
  jk_z\tilde\rho_c\left(\tilde\Phi_i^+e^{-j\Omega_pt'}
    +\tilde\Phi_i^-e^{j\Omega_pt'}\right)dx\,d^2\bs  
  k.  
\end{equation} 
The expression of $F_z^{(1)}$ is formally similar to the expression of
$F_y^{(1)}$ and can be transformed in a similar manner, using Eq.~\eqref{eq:76}:
\begin{equation}
  \label{eq:144}
    F_z^{(1)}=\frac{I}{2\pi^2}\iiint\Im[\tilde\rho_i^+\exp(j\Omega_p 
  t')]k_z\tilde\Phi_cdx\,d^2\bs k. 
\end{equation}
where we have $\tilde\rho_i^R\equiv\Re(\tilde\rho_i^+)$ and
$\tilde\rho_i^I\equiv\Im(\tilde\rho_i^+)$. Noting that
$\tilde\Phi_i^-=-\tilde\Phi_i^+=k_z\tilde\Phi_c/2$, it is
straightforward to recast Eq.~\eqref{eq:133} as:
\begin{equation}
  \label{eq:145}
  F_z^{(2)}=\frac{I}{4\pi^2}\iiint\tilde\rho_ck_z^2\tilde\Phi_c\sin(\Omega_pt')dx\,d^2\bs k.
\end{equation}
Given the parities in $x$ of the functions $\tilde\Phi_c$,
$\tilde\rho_c^R$ and $\tilde\rho_c^I$, the imaginary part of
$\tilde\rho_c$ yields a null contribution in the integral above, so
that $\tilde\rho_c$ in the integrand can be replaced by
$\tilde\rho_c^R$.
We can now write, using Eqs.~\eqref{eq:144} and~\eqref{eq:145}:
\begin{equation}
  \label{eq:146}
F_z = F_z^C\cos \Omega_p t' + F_z^S\sin\Omega_p t',
\end{equation}
with:
\begin{equation}
\label{eq:147}
F_z^C = \frac{I}{2\pi^2} \iiint \tilde\rho_i^I k_z \tilde\Phi_c
dx\,d^2\bs k
\end{equation}
and
\begin{equation}
\label{eq:148}
F_z^S = \frac{I}{2\pi^2} \iiint k_z \tilde\Phi_c \left( \tilde\rho_i^R
  + \frac{\tilde\rho_c^R k_z}{2}\right)  dx\,d^2\bs k.
\end{equation}

\subsection{Reduction of the force components to a non-dimensional
  form}
\label{sec:reduct-force-comp}
In this section we transform the six integrals that define the cosine
and sine amplitude of the three force components.  Following
\citet{2017MNRAS.472.4204M}, we define:
\begin{equation}
  \label{eq:149}
  k_c=\sqrt\frac{3\Omega_p\gamma}{2\chi}
\end{equation}
and the dimensionless form of the wavevectors $k_y$ and $k_z$:
\begin{equation}
  \label{eq:150}
  K_y=k_y/k_c\mbox{~~~and~~~}K_z=k_z/k_c.
\end{equation}
We furthermore define:
\begin{equation}
  \label{eq:151}
  K_0=(K_y^2+K_z^2)^{1/2}.
\end{equation}
Note that $K_0$ differs from $K$ defined at Eq.~\eqref{eq:78}: we have
the relationship
\begin{equation}
  \label{eq:152}
  K_0^3=K_yK.
\end{equation}
In addition to the non-dimensional
functions respectively solutions of the Eqs.~\eqref{eq:93}
and~\eqref{eq:94}, we shall also need the solution of the circular
problem, which is:
\begin{equation}
  \label{eq:153}
  \tilde\rho_c=s(R_K+jI_K),
\end{equation}
where $R_K$ and $I_K$ are solutions of the system of differential
equations~(75) and~(76) of \citet{2017MNRAS.472.4204M}, and where:
\begin{equation}
  \label{eq:154}
  s=-\frac{2(\gamma-1)k^2(L-L_c)}{3\Omega_pk_yc_s^2}.
\end{equation}
We mention that $R_K+jI_K$ is the particular solution $T_{X_c,K}$ with
$X_c=0$.

\begin{figure*}
\includegraphics[width=6cm]{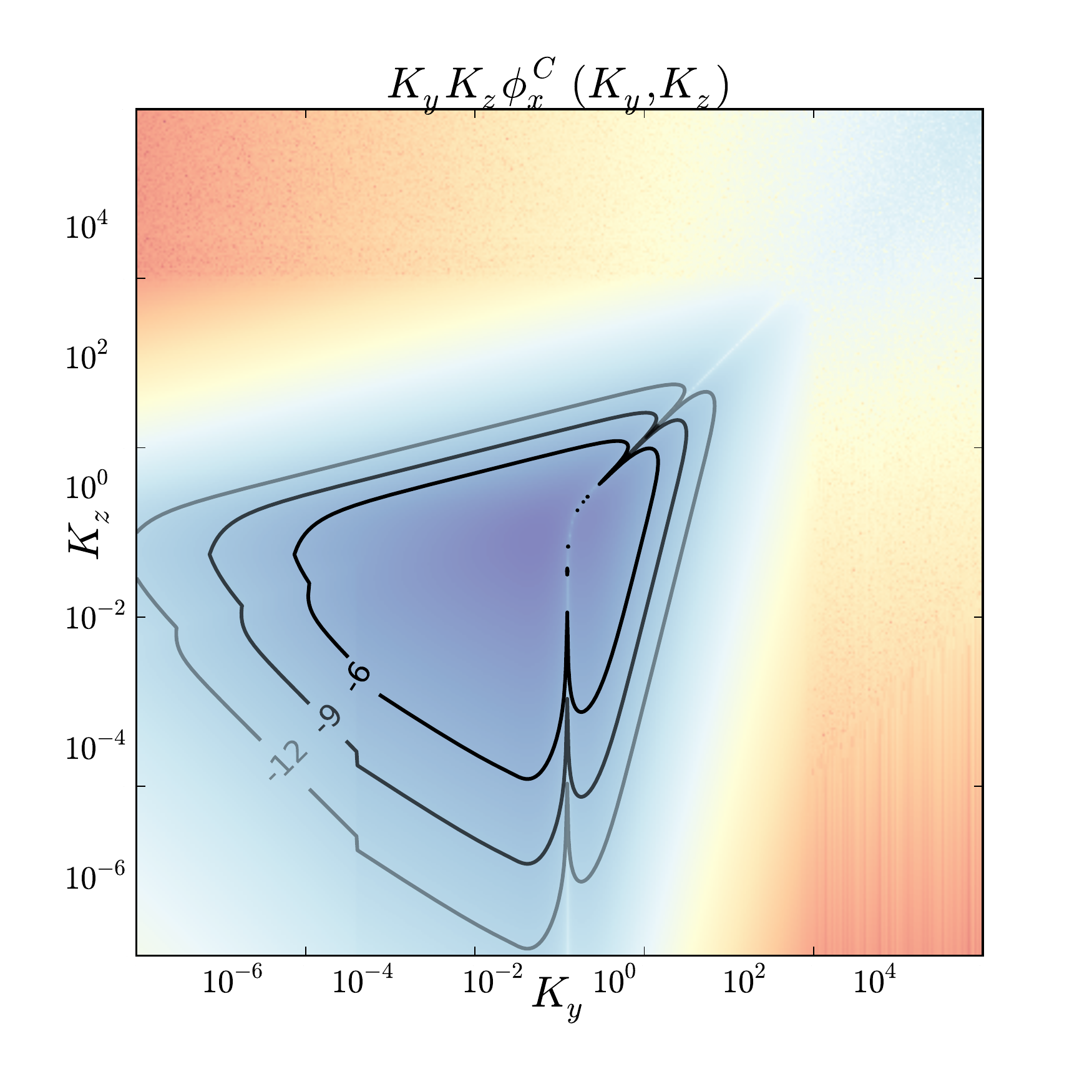}\includegraphics[width=6cm]{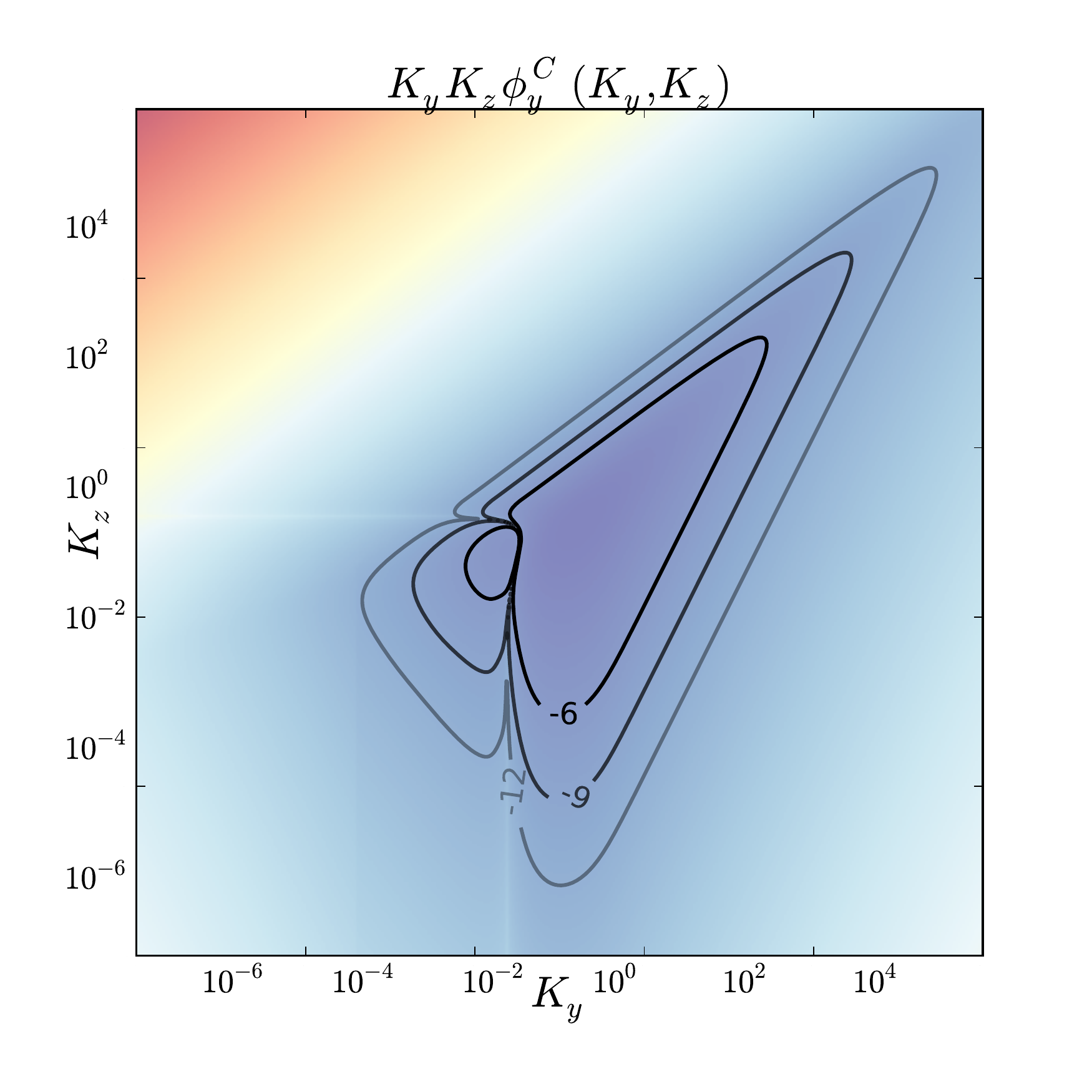}\includegraphics[width=6cm]{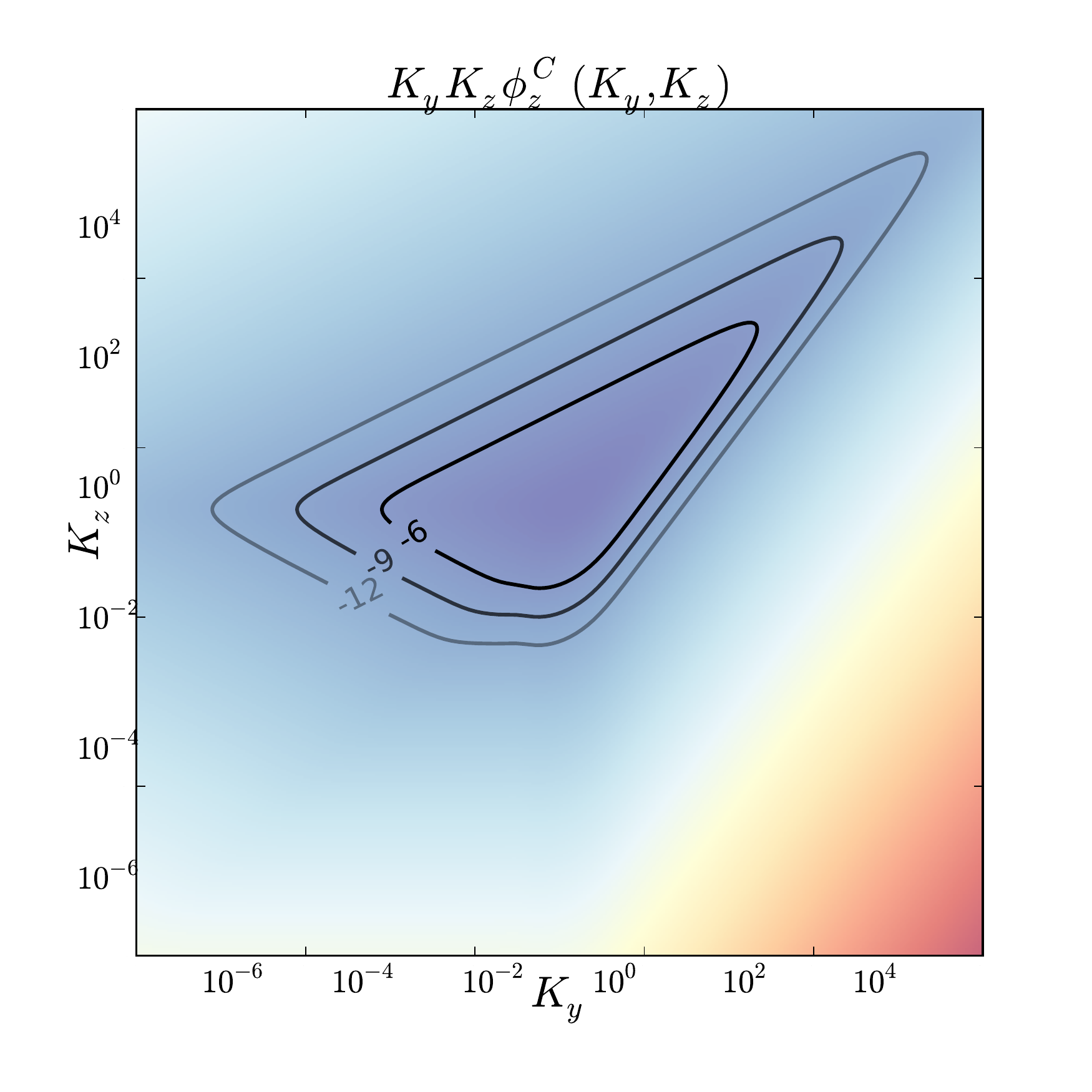}\\
\includegraphics[width=6cm]{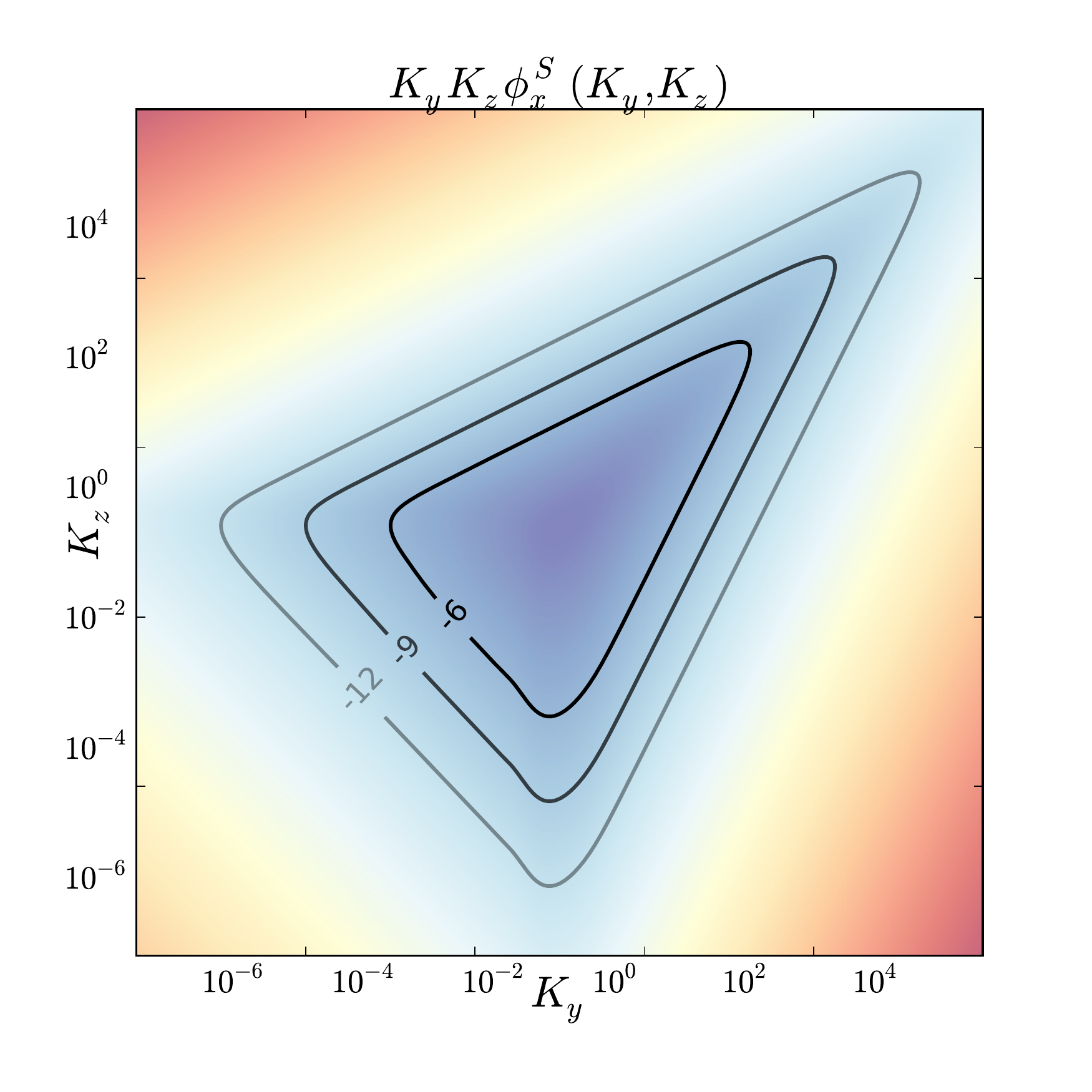}\includegraphics[width=6cm]{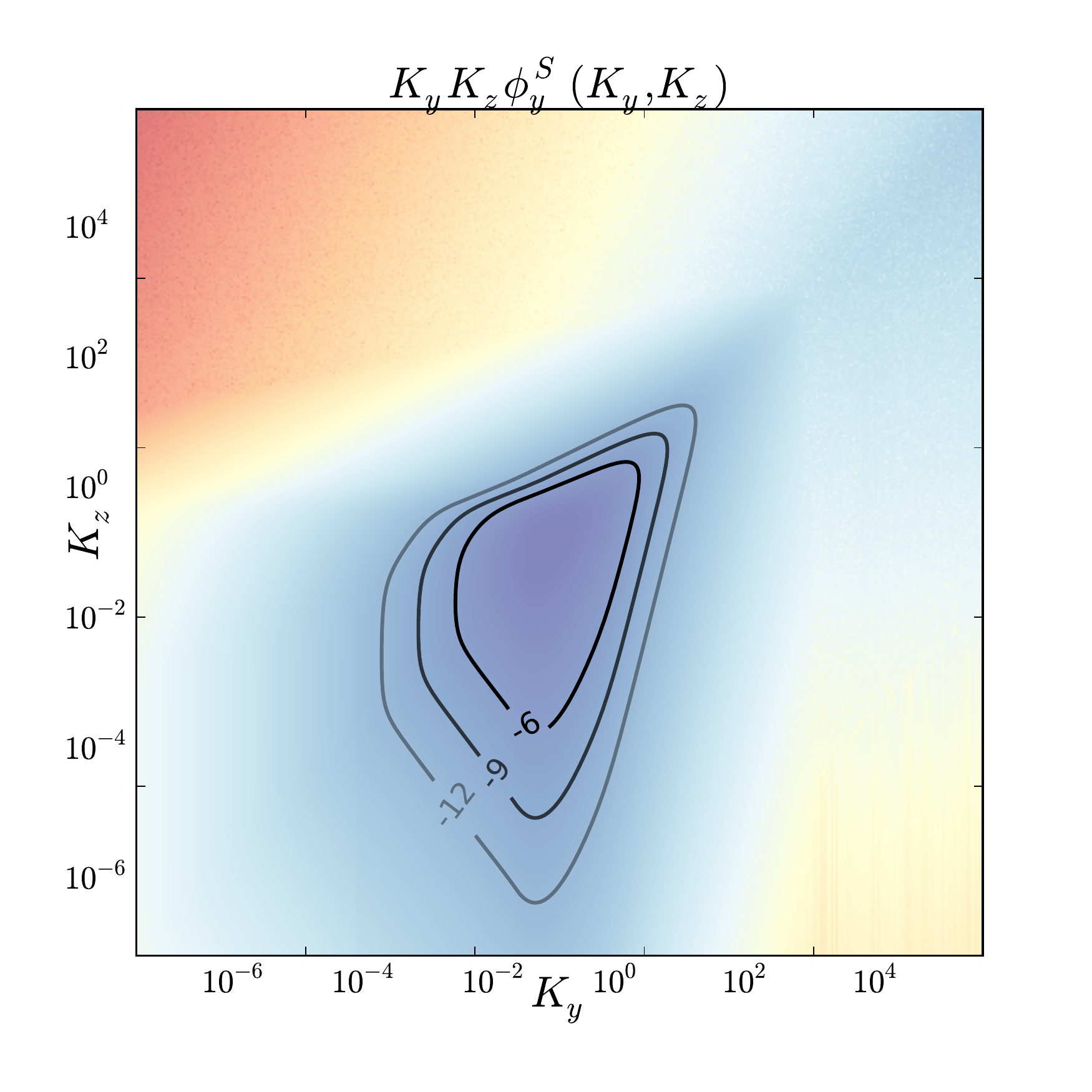}\includegraphics[width=6cm]{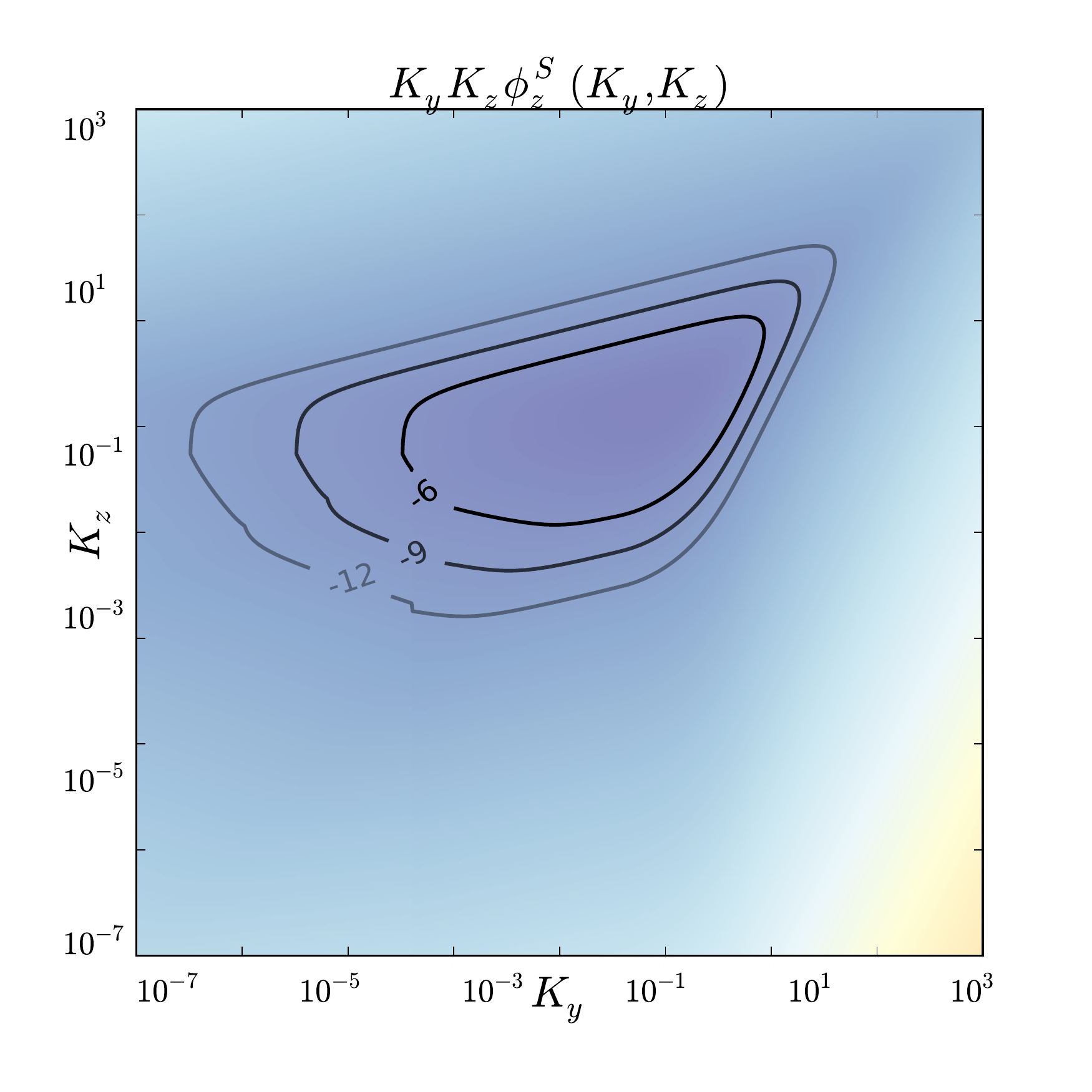}
\caption{Colour maps of the integrand function of $K_y$ and $K_z$ for
  the evaluation of the sine (lower row) and cosine (upper row)
  components of the radial $(f_x^c,f_x^s)$, azimuthal $(f_y^c,f_y^s)$
  and vertical $(f_z^c,f_z^s)$ forces, from left to right. The
  values are normalized to the peak value. We add iso-contours at
  $10^{-6}$, $10^{-9}$ and $10^{-12}$. }
\label{fig:forces}
\end{figure*}

Using Eqs.~\eqref{eq:81},
\eqref{eq:84}, \eqref{eq:129}, \eqref{eq:149}, \eqref{eq:150}
to~\eqref{eq:154}, we can write:
\begin{equation}
  \label{eq:155}
  F_x^C=\frac{GMae(\gamma-1)(L-L_c)k_c^3}{3\pi\Omega_pc_s^2}
  f_x^C
\end{equation}
where
\begin{equation}
  \label{eq:156}
  f_x^C=\iiint\left(2K_0T^R+\frac{K_0^2}{K_y}R_K'\right)e^{-|X|}\mathrm{sgn}(X)dX\,d^2\bs
  K
\end{equation}
is a dimensionless constant, and where for conciseness we note
$T^R\equiv\Re(T_{X_c,K})$.  Eq.~\eqref{eq:155} can be recast as:
\begin{equation}
  \label{eq:157}
  F_x^C=F_0ef_x^C
\end{equation}
with
\begin{equation}
  \label{eq:158}
  F_0=\frac{\gamma^{3/2}(\gamma-1)GMa(L-L_c)(3\Omega_p/2)^{1/2}}{2\pi c_s^2\chi^{3/2}}.
\end{equation}
Similarly, we have:
\begin{eqnarray}
  \label{eq:159}
  F_x^S&=&F_0ef_x^S\\
  \label{eq:160}
  F_y^C&=&F_0ef_y^C\\
  \label{eq:161}
  F_y^S&=&F_0ef_y^S\\
  \label{eq:162}
  F_z^C&=&F_0if_z^C\\
  \label{eq:163}
  F_z^S&=&F_0if_z^S
\end{eqnarray}
where
\begin{eqnarray}
  \label{eq:164}
  f_x^S&=&-\iiint 2K_0(T^I-I_K)e^{-|X|}\mathrm{sgn}(X)dX \,d^2\bs K\\
  \label{eq:165}
  f_y^C&=&-\iiint (2K_yT^I+K_0I_K')e^{-|X|}dX \,d^2\bs K\\
  \label{eq:166}
  f_y^S&=&-\iiint 2K_y(T^R-R_K)e^{-|X|}dX \,d^2\bs K\\
  \label{eq:167}
  f_z^C&=&-\iiint W^I\frac{K_z^2}{K_y}e^{-|X|}dX \,d^2\bs K\\
  \label{eq:168}
  f_z^S&=&-\iiint \frac{K_z^2}{K_y}(W^R-R_K)e^{-|X|}dX \,d^2\bs K
\end{eqnarray}
are dimensionless constants. We determine these constants as
follows. For definiteness, we specify hereafter to the case of
$f_x^C$, the other five constants being obtained in a strictly similar
fashion. We define:
\begin{equation}
  \label{eq:169}
  \phi_x^C(K_y,K_z)=\int_{-\infty}^{+\infty}\left(2K_0T_{X_c,K}^R+\frac{K_0^2}{K_y}R_K'\right)e^{-|X|}\mathrm{sgn}(X)dX.
\end{equation}
The integrand of this expression is the same as that of
Eq.~\eqref{eq:156}, except that we have explicitly written the
dependence of $T$ on $X_c$ and $K$, and the integration is performed
over $X$ only. The different functions that feature in the integrand
of this expression are obtained following the method of
section~\ref{sec:resolution-practice}. The integral is a function of
$K_y$ and $K_z$ only. Indeed $K_0$ (Eq.~\ref{eq:151}), $X_c=2K_0/3K_y$
(Eqs.~\ref{eq:77} and~\ref{eq:79}) and $K$ (Eq.~\ref{eq:152}) are
functions of $K_y$ and $K_z$ only.  Once the functions of the
integrand are evaluated, we perform the integral of Eq.~\eqref{eq:169}
using Simpson's method. The dimensionless coefficient $f_x^C$ is then
obtained by evaluating:
\begin{equation}
  \label{eq:170}
  f_x^C=\iint \phi_x^C(K_y,K_z)dK_y\,dK_z.
\end{equation}
We can recast this expression as:
\begin{equation}
  \label{eq:171}
  f_x^C=4\int_0^{+\infty}\int_0^{+\infty}K_yK_z
  \phi_x^C(K_y,K_z)d\log K_y\,d\log K_z.
\end{equation}
This, in practice, corresponds to the way we numerically evaluate this
integral. We use a 2D Simpson method over the plane
$(K_y,K_z)$, with $(K_y,K_z)\in[10^{-7},10^6]\times [10^{-7},10^6]$,
and with a constant spacing in $\log K_y$ and $\log K_z$. We check
that our domain of integration is sufficiently large that the
integrand $K_yK_z\phi_x^C(K_y,K_z)$ is negligible near its edges (see
Fig.~\ref{fig:forces}). We apply the same reasoning to all the
components and we obtain:
\begin{eqnarray}
f_x^C &=&  -0.507 \label{eq:172}\\
f_x^S &= & 1.440  \label{eq:173}\\
f_y^C & = & 0.737  \label{eq:174}\\
f_y^S  &= &0.212  \label{eq:175}\\
f_z^C &=& 1.160  \label{eq:176}\\
f_z^S &=& 0.646  \label{eq:177}
%f_x^C &=&  -0.506925551....
%f_x^S &= & 1.44007542.....
%f_y^C & = & 0.737379261....
%f_y^S  &= &0.211504.....
%f_z^C = 1.159868154....
%f_z^S = 0.64601174....
\end{eqnarray}
We show in Fig.~\ref{fig:forces}) that we consider at least all
the area embedding the contributions higher than $10^{-12}$ times the
peak value. Moreover, we have checked that these values are converged
to at least four significant digits with respect to the resolution in
$K_y$ and $K_z$, with respect to the resolution with which we solve
the differential equations in $X$ for $T$ and $W$, and with respect to
the value of $X_0$ that we use to find these functions (see
section~\ref{sec:resolution-practice}).

\section{Time evolution of eccentricity and inclination}
\label{sec:eccentr-incl-growth}
Having worked out the time-dependent force acting on the planet, we
can now evaluate the time derivatives of the eccentricity and
inclination. Their expressions are respectively \citep[][and
refs. therein]{2004ApJ...602..388T}:
\begin{equation}
  \label{eq:178}
  \dot e=\frac{1}{M\Omega_p a}[F_x\sin(\Omega_p t)+2F_y\cos(\Omega_p
  t)]
\end{equation}
and:
\begin{equation}
  \label{eq:179}
  \dot i=\frac{1}{M\Omega_p a}F_z\cos(\Omega_p t'),
\end{equation}
which, upon averaging over one orbital period, read:
\begin{equation}
  \label{eq:180}
  \bar{\dot e}=\frac{F_0e}{M\Omega_p a}\left(\frac{f_x^S}{2}+f_y^C\right)
\end{equation}
and
\begin{equation}
  \label{eq:181}
  \bar{\dot i}=\frac{F_0i}{M\Omega_p a}\frac{f_z^C}{2}.
\end{equation}
Defining the thermal time $t_\mathrm{thermal}$ as:
\begin{equation}
  \label{eq:182}
  t_\mathrm{thermal}=\frac{M\Omega_pa}{F_{00}}=\frac{c_s^2\Omega_p\lambda_c}{2(\gamma-1)G^2M\rho_0},
\end{equation}
where $F_{00}$ is the value of $F_0$ when the planet is non-luminous
($L=0$), and where
\begin{equation}
  \label{eq:183}
  \lambda_c={k_c}^{-1}, 
\end{equation}
we have:
\begin{eqnarray}
  \label{eq:184}
  \frac{\bar{\dot e}}{e}&=&\frac{1.46}{t_\mathrm{thermal}}(\ell-1)\\
  \label{eq:185}
  \frac{\bar{\dot i}}{i}&=&\frac{0.58}{t_\mathrm{thermal}}(\ell-1),
\end{eqnarray}
with
\begin{equation}
  \label{eq:186}
  \ell=\frac{L}{L_c}.
\end{equation}
It is instructive to compare $t_\mathrm{thermal}$ to the
damping time $t_\mathrm{wave}$ defined by \citet{2004ApJ...602..388T},
which reads, with our notation:
\begin{equation}
  \label{eq:187}
  t_\mathrm{wave}=\left(\frac{M}{M_\star}\right)^{-1}\left(\frac{\sqrt{2\pi}H\rho_0
      a^2}{M_\star}\right)^{-1}\left(\frac{c_s}{a\Omega_p}\right)^4\Omega_p^{-1}.
\end{equation}
We find:
\begin{equation}
  \label{eq:188}
  \frac{t_\mathrm{thermal}}{t_\mathrm{wave}}=\sqrt\frac\pi
  2\frac{1}{\gamma(\gamma-1)}\frac{\lambda_c}{H}\approx 2.24 \frac{\lambda_c}{H},
\end{equation}
where the approximation corresponds to the particular case
$\gamma=1.4$. The dimensionless parameter $\ell$ is zero when the
planet is non-luminous. In that case, we see that the eccentricity and
inclination of the planet are damped, on a timescale that is much
shorter than that of \citet{2004ApJ...602..388T}. Namely, we have in
that case:
\begin{equation}
  \label{eq:189}
  \left.\frac{\bar{\dot e}}{e}\right|_\mathrm{thermal}=0.84\left(\frac{H}{\lambda_c}\right)\left.\frac{\bar{\dot e}}{e}\right|_\mathrm{TW04}
\end{equation} 
and
\begin{equation}
  \label{eq:190}
  \left.\frac{\bar{\dot i}}{i}\right|_\mathrm{thermal}=0.48\left(\frac{H}{\lambda_c}\right)\left.\frac{\bar{\dot i}}{i}\right|_\mathrm{TW04}.
\end{equation}
Typically, $H/\lambda_c\sim 10$ at a few astronomical units in a
protoplanetary disc around a solar-type star
\citep{2017MNRAS.472.4204M}. This implies that the eccentricity and
inclination damping on a non-luminous embryo is completely dominated
by thermal effects, which are nearly one order of magnitude more
important than those arising from wave launching. This strong damping
is in agreement with the findings of \citet{2017arXiv170401931E} who
found that the eccentricity and inclination of a non-luminous low-mass
planet embedded in a radiative disc was indeed much stronger than that
expected from \citeauthor{2004ApJ...602..388T}'s formulae. When
$\ell >1$, i.e. when the planet's luminosity is larger than the
critical luminosity $L_c$, Eqs.~\eqref{eq:184} and~\eqref{eq:185} show
that the eccentricity and inclination grow exponentially. When $\ell$
is not too close to one, thermal effects largely dominate over those
arising from wave launching, and the eccentricity grows at a rate
$1.46/0.58\approx 2.5$ times larger than the inclination. This is in
correct agreement with the findings of \citet{2017arXiv170401931E}. We
note that $\ell$ can be of order of a few for an Earth-mass embryo
with a mass doubling time of the order of $10^5$~yr in a typical
protoplanetary disc at a few astronomical units. Naturally, the
parameter space is so large that significantly more work is required
to assess the relative importance of thermal effects over wave
effects. Such study is largely beyond the scope of this work,
but we anticipate that, in general, thermal effects should be dominant
for planetary embryos in planet-forming regions of protoplanetary discs.

\section{Discussion}
\label{sec:discussion}
In the following we discuss a few aspects of our findings, namely
their relation with the non-linear corotation torque and horseshoe
dynamics, and how they compare to the results previously obtained for
planets in circular orbits and perturbers in non-sheared, homogeneous
media. We also compare the different timescales for migration,
eccentricity and inclination evolution under either thermal or wave-supported
disturbances. We discuss the different regimes of eccentricity
(opposing shear-dominated to headwind-dominated cases) and we finally
briefly discuss the mass range over which the present results are
expected to apply.

\subsection{Relation to horseshoe dynamics}
\label{sec:relat-hors-dynam}
The processes that we have presented here, which are captured by a
linear analysis, occur in the vicinity of the orbit, a location that
also gives rise to the corotation torque.  It is well known that, in
an inviscid disc, the corotation torque always becomes non-linear upon
a timescale that depends on the planetary mass
\citep{2009arXiv0901.2265P}. However, the horseshoe dynamics should
have little impact, if any, on the processes presented here. The
response time of the thermal force, which is the time required for the
heat released by the planet to diffuse over a lengthscale $\lambda$,
is of the order of the dynamical timescale $\Omega_p^{-1}$. This
timescale is much shorter than the timescale for the horseshoe U-turns
of low-mass planets, which can amount to tens of orbits \citep[see
e.g.][]{2017MNRAS.472.4204M}. Note also that the horseshoe motion
corresponds to minute perturbations of the unperturbed sheared
flow. We have seen in section~\ref{sec:simpl-main-equat} that the
perturbations of this flow can be neglected: the effects that we
present here are described by an advection-diffusion equation, where
the advection part comes from the unperturbed flow and the diffusion
occurs in the vicinity of a point-like source that describes an
epicycle. A full description of the dynamics of the coorbital regions
that includes simultaneously thermal effects and the non-linear
horseshoe dynamics is probably too complex to be tackled analytically
and should be undertaken through numerical experiments. It is however
reasonable to expect that such experiments would confirm that thermal
effects are essentially decoupled from the horseshoe dynamics, for the
following reasons:
\begin{itemize}
\item The numerical experiments of \citet{2017arXiv170401931E},
 which were performed prior to the existence of this work and
 were therefore not aimed at checking analytical predictions with a
 high accuracy, do show that planets are subjected to effects similar
 to those described here (i.e. excitation of eccentricity and
 inclination at large luminosity, as opposed to a strong damping of
 these quantities when they are non-luminous). The planetary masses
 considered in these experiments range from a fraction of an Earth
 mass to about ten Earth masses. The horseshoe region is resolved, and
 the horseshoe dynamics captured, for planets above one Earth-mass in
 those experiments\footnote{In these experiments the horseshoe region
   of a one Earth-mass planet spans radially 11~cells.}.
\item Arguments similar to those exposed above can be applied to the
  damping of eccentricity that occurs as a consequence of
  wave-launching. In that case it is primarily due to wave excitation
  at coorbital Lindblad resonances \citep{arty93b}. Those share their
  location with the horseshoe region. Yet, there is again a large
  difference between the response time of the waves (which is
  typically the dynamical timescale $\Omega_p^{-1}$) and the horseshoe
  U-turn time so that one may expect that
  the eccentricity and inclination damping may be quite insensitive to
  the horseshoe dynamics. There is a large body of numerical studies
  of the eccentricity damping in isothermal discs that support this
  view, as they confirm the analytical predictions of
  \citet{2004ApJ...602..388T}, even for planets subjected to a
  horseshoe drag \citep[e.g.][]{2007A&A...473..329C}.
\end{itemize}
We finally mention that planets that have a luminosity $L$ larger than
the critical luminosity $L_c$ experience a growth of eccentricity and
inclination. In these circumstances, the coorbital corotation torque
is quenched \citep{2012MNRAS.419.2737H}.

\subsection{Comparison to earlier results}
\label{sec:comp-earl-results}
\begin{figure}
  \centering
  \includegraphics[width=\columnwidth]{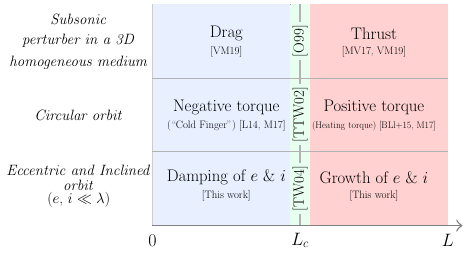}
  \caption{\label{fig:colortable}Summary of the three cases mentioned
    in the text. For reasons of consistency we use a colour code, in
    the electronic version, similar to that of
    \citet[][{[VM19]}]{2019MNRAS.483.4383V}. The cases for which the
    luminosity feedback dominates ($L>L_c$) are represented on a red
    background, whereas those for which the luminosity is sub-critical
    are represented with a blue background. For $L\approx L_c$, one
    recovers the adiabatic situation (represented on a green
    background).}
\end{figure}

The analysis presented here is the last one of a series of three
analytical studies devoted to the role of thermal diffusion and
luminosity feedback in situations of interest for planet-disc
interaction.
\begin{itemize}
\item The first one dealt with the simplest setup, that of a perturber
  moving across a 3D medium at rest (a setup that gives
  rise to dynamical friction when the luminosity feedback is not taken
  into account), with special emphasis on the low Mach number
  regime. In that case, the two additional forces (the force arising from
  the perturber's luminosity and that arising from the mere inclusion
  of thermal diffusion, when the perturber is non-luminous) have been
  studied in two separate publications. The former, dubbed heating
  force, has been studied by \citet[][{[MV17]}]{2017MNRAS.465.3175M},
  while the latter, named cold thermal force, has been evaluated
  analytically by \citet[][{[VM19]}]{2019MNRAS.483.4383V} and
  corroborated numerically in that same work. The net force arising from
  thermal effects is a drag when the perturber's luminosity $L$ is
  smaller than a critical luminosity $L_c$ that has the exact same
  expression as Eq.~\eqref{eq:58}, and a thrust otherwise.
\item The second analysis dealt with planets on circular orbits, and
  analysed the new torque components that arise from the heat release
  by the planet (or heating torque) and from the inclusion of thermal
  diffusion (or cold thermal torque). While primarily intended as an
  analytical follow-up of the work of
  \citet[][{[BLl+15]}]{2015Natur.520...63B}, this analysis allowed to
  identify the effect found by
  \citet[][{[L+14]}]{2014MNRAS.440..683L}, and dubbed the \emph{cold
    finger effect} by these authors, with the cold thermal
  torque. Again, this analysis showed that the net thermal torque
  changes sign for the perturber's luminosity $L=L_c$, which has again
  the exact same value as that given by Eq.~\eqref{eq:58}.
\end{itemize}
These two cases and the case presented in this work are summarized in
Fig.~\ref{fig:colortable}. The critical luminosity $L_c$ of
Eq.~\eqref{eq:58} appears as a universal watershed at which the net
thermal effects change sign. This has been discussed by
\citet[][section~5.7 and Fig.~3]{2017MNRAS.472.4204M}. In an adiabatic
case, an enthalpy (or temperature) peak surrounds the planet (for
low-mass planets the peak is nearly the negative of the potential
well). The introduction of thermal diffusion tends to reduce large
temperature gradients and flattens this peak. The release of energy
into the surrounding gas by a luminous perturber rebuilds the
peak. When the luminosity of a perturber with low Mach number is equal
to the critical luminosity $L_c$, the temperature peak that surrounds
the perturber has same amplitude and shape as that of the adiabatic
case (even though the underlying physical processes responsible for
the peak are quite different) and the thermal effects cancel out.
For this specific case with $L=L_c$, one recovers in an unsheared
homogeneous gas the force studied by
\citet[][{[O99]}]{1999ApJ...513..252O}, whereas for the cases of a
planet on circular orbit and the more general case with finite but
small eccentricity and inclination, one
recovers\footnote{\label{foo:iso}The studies of TTW02 and TW04 are
  strictly speaking for isothermal discs, rather than adiabatic
  discs. None the less, one can infer from these works the behaviour in
  an adiabatic disc by substituting the isothermal sound speed with
  the adiabatic sound speed.} respectively the torque of
\citet[][{[TTW02]}]{tanaka2002} and the damping rates of
\citet[][{[TW04]}]{2004ApJ...602..388T}. Due to the large value of
thermal forces on low-mass planets in planet-forming regions of
protoplanetary discs, only when $L\approx L_c$ do we nearly recover
the results for adiabatic discs. The vertical green band at $L_c$ (in
the electronic version) has been intentionally represented quite
narrow to illustrate this effect.

\subsection{Comparison of timescales}
\label{sec:comp-timesc}
\begin{figure}
  \centering 
  \includegraphics[width=\columnwidth]{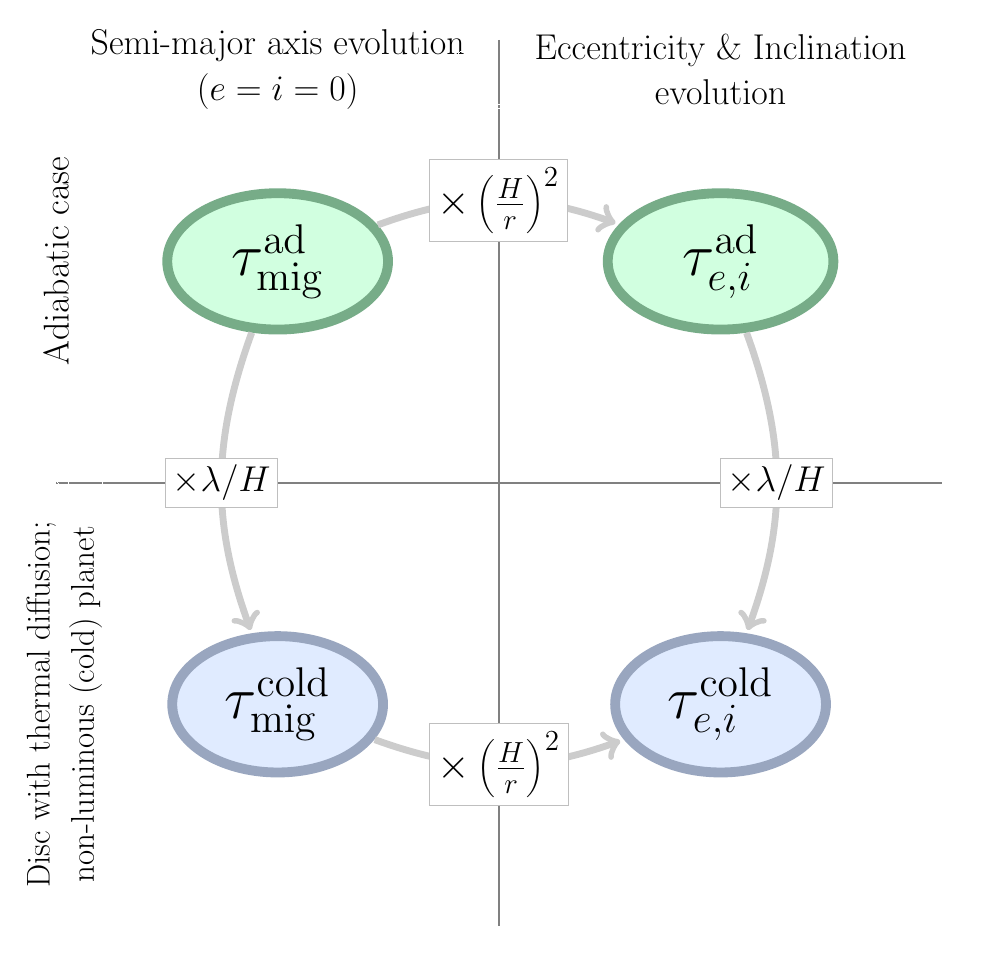}
  \caption{\label{fig:diag}Order of magnitude relationships between
    different timescales. The left column shows the semi-major axis
    variation timescale (migration time) and the right column the
    damping timescale of eccentricity and inclination. The top row
    corresponds to planets in adiabatic discs, while the bottom row
    corresponds to non-luminous planets in discs with thermal
    diffusion.}
\end{figure}

The effects presented here are considerable, and dominate over those
arising from wave-launching for low-mass planets except when
$L\approx L_c$. Eqs.~\eqref{eq:189} and \eqref{eq:190} show that the
eccentricity and inclination damping timescales of a non-luminous
planet are a factor of $\lambda/H$ shorter than those given by
\citet{2004ApJ...602..388T}. We note that the same factor appeared
between the migration timescale arising from the cold thermal torque
and that arising from wave-launching
\citep[][{Eq.~[137]}]{2017MNRAS.472.4204M}. The timescale for
eccentricity and inclination damping being, in adiabatic discs, a
factor of $(H/r)^2$ shorter than the migration timescale
\citet{arty93b, 2004ApJ...602..388T}, we therefore expect the same
ratio to hold between the timescales for the evolution of the
eccentricity and inclination and that for the evolution of the
semi-major axis, under thermal forces. From Eqs.~(133), (138)
and~(145) of \citet{2017MNRAS.472.4204M}, we see that the migration
timescale of a non-luminous planet subjected to the thermal torque is,
in order of magnitude:
\begin{equation}
  \label{eq:191}
  t_\mathrm{mig}\sim\frac{a^2\Omega_p^3\lambda_c}{G^2M_p\rho_0},
\end{equation}
which is indeed a factor $\sim (H/r)^{-2}$ larger than the thermal
time of Eq.~\eqref{eq:182}. We summarize these different relationships
in Fig.~\ref{fig:diag}.
Note that the estimates of the timescales from
thermal forces are for a non-luminous planet. These characteristic
timescales for a planet with a luminosity largely in excess of $L_c$
would be even shorter\footnote{They would then be \emph{growth}
  timescales for the eccentricity and inclination, and a timescale of
  \emph{outward} migration for the semi-major axis.}.

\subsection{Different regimes of eccentricity}
\label{sec:diff-regim-eccentr}
\begin{figure*}
  \centering
  \includegraphics[width=\textwidth]{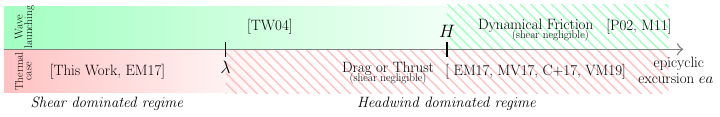}
  \caption{\label{fig:ecc}The different regimes of eccentricity
    mentioned in the text. The upper part refers to work on
    eccentricity damping due to wave-launching in adiabatic discs (or
    in isothermal discs, see footnote~\ref{foo:iso}), while the lower
    part refers to work on the eccentricity damping or growth under
    thermal disturbances. The hatched areas show the regimes where the
    time evolution of the eccentricity is described by a dynamical
    friction calculation, while those with a gradient background
    depict the shear-dominated regimes. The numerical experiments of
    \citet{2017arXiv170401931E} cover the two regimes for thermal
    disturbances.}
\end{figure*}

Planets with super-critical luminosity ($L>L_c$) experience an
exponential growth of eccentricity and inclination with time. At some
point the hypothesis of Eq.~\eqref{eq:14} that the epicyclic and
vertical excursions $E$ and $I$ are much smaller than the size of the
thermal disturbance ceases to be valid. It is straightforward to
realise that in these circumstances the thermal disturbance tends
towards that triggered by a perturber in an unsheared medium. The
velocity $v_p $ of the perturber with respect to the ambient gas is
indeed larger than $\lambda_c\Omega_p$. The response time of the
thermal disturbance is then \citep{2017MNRAS.465.3175M}:
\begin{equation}
  \label{eq:192}
\tau\sim \chi/v_p^2\ll\chi/(\Omega_p^2\lambda_c^2)=\Omega_p^{-1}
\end{equation}
The response time being shorter than the shear timescale, the shear
becomes unimportant and the thermal force tends towards that of
unsheared media \citep{2017MNRAS.465.3175M,2019MNRAS.483.4383V}. This
regime has been named the headwind-dominated regime by
\citet[][section~4.5]{2017arXiv170401931E}, as opposed to the regime
of low eccentricities and inclinations that we studied in the present
work, which was referred to as the shear-dominated regime. In the same
vein, this effect has been named the \emph{hot trail effect} by
\citet[][{[C+17]}]{2017arXiv170606329C}. The same kind of transition
from shear-dominated to headwind-dominated 
occurs for the pressure-supported wake, except that it occurs at
epicyclic or vertical excursions comparable to the pressure length
scale $H$. The calculation of the eccentricity and inclination damping
rates of \citet{2004ApJ...602..388T} requires that $e \ll H/r$ and
$i \ll H/r$, and this, in general, is the case of estimates of damping
rates based on a sum of resonances \citep{WW88,arty93b}, as for
eccentricities larger than the disc's aspect ratio the use of series
on resonances becomes impractical \citep{paplar2000}. When the
eccentricities are larger than the disc's aspect ratio, a dynamical
friction calculation is much more convenient \citep[][resp. {[P02]}
and {[M11]}]{2002A&A...388..615P,2011ApJ...737...37M}. The case of
inclinations larger than the disc's aspect ratio is slightly
different, as the planet spends a fraction of its orbit outside of the
disc, but is also conveniently dealt with using a dynamical friction
calculation \citep{2012MNRAS.422.3611R}. We depict the different
regimes of eccentricity in Fig.~\ref{fig:ecc}.
The asymptotic values reached by the eccentricity in the numerical
experiments of \citet{2017arXiv170401931E} typically fall within the
$[\lambda,H]$ interval. In these conditions the equilibrium
eccentricity is given by the balance between the time-varying force of
\citet{2004ApJ...602..388T} for the pressure-supported disturbance,
and by the heating force in an unsheared medium
\citep{2017MNRAS.465.3175M,2019MNRAS.483.4383V}.

\subsection{Dependence of thermal forces on the perturber's mass}
\label{sec:depend-therm-forc}
As discussed by \citet{2017MNRAS.465.3175M} in the context of
unsheared media, the thermal forces are expected to suffer a decay
with respect to their linearly predicted value when $M_p>M_c$, where the
critical mass $M_c$ is given by:
\begin{equation}
  \label{eq:193}
  M_c=\frac{\chi c_s}{G}.
\end{equation}
When the perturber's mass is much smaller than $M_c$, the heat
diffusion time across the planetary Bondi radius is much smaller than
the acoustic time across the Bondi radius, which guarantees that the
energy released by the planet in its immediate neighbourhood reaches
outside the Bondi sphere, where our linear analysis is valid,
as an excess of internal energy. This no longer needs to be the case
when $M_p\gtrsim M_c$, however. Using arguments based on the yield of
the heating force, \citet{2017MNRAS.465.3175M} argue that the latter
has to be cut off for $M_p>M_c$. The numerical value of $M_c$, in
planet-forming regions of protoplanetary discs, is of the order of an
Earth-mass, although this value can vary greatly as a function of the
position in the disc and as the discs evolves and cools
\citep{2017MNRAS.465.3175M,2017MNRAS.472.4204M}. The numerical
experiments of \citet{2014MNRAS.440..683L},
\citet{2015Natur.520...63B} and~\citet{2017arXiv170401931E} are all
compatible with a cut-off of thermal effects above masses
commensurable with an Earth-mass. An accurate determination of the
cut-off law probably requires high-resolution calculations that
resolve the Bondi sphere, as this effect cannot be captured by means
of a linear calculation. We mention none the less that thermal effects
are so vigorous that even in numerical experiments of
\citet{2017arXiv170401931E}, a sizeable impact of the planet's
radiative feedback on its eccentricity and inclination is found up to
approximately 5-10 Earth masses.

\section{Conclusions}
\label{sec:conclusion}
We have worked out the time-dependent force acting on a planetary
embryo embedded in a gaseous protoplanetary disc, using linear
perturbation theory, when thermal diffusion in the gas is taken into
account, with or without heat release by the planet into the
surrounding gas.

We find that this thermal force has a strong impact on the
eccentricity and inclination of the embryo, with an outcome that
depends on the embryo's luminosity~$L$. When the latter is smaller
than the critical luminosity $L_c$ defined at Eq.~\eqref{eq:58}, the
thermal force leads to a damping of eccentricity and inclination,
typically stronger by up to an order of magnitude than the damping due
to wave-launching considered so far \citep{WW88,arty93b,
  1994ApJ...423..581A, 2004ApJ...602..388T}.  The sign of thermal
forces reverses when the embryo's luminosity is $L_c$. Only in
the regime where $L\approx L_c$ does the damping due to wave-launching
play a role, as it is otherwise masked by the strong effect of the
thermal force. For luminosities significantly greater than $L_c$, as
can be expected for Earth-like embryos with mass doubling times
shorter than $100$~kyr at a few astronomical units in discs similar
to the Minimum Mass Solar Nebula, the eccentricity and inclination
grow exponentially over a short time scale of the order of hundreds of
orbits only. The outcome of such growth has been studied numerically
by \citet{2017arXiv170401931E} and \citet{2017arXiv170606329C} (in
2D discs). The critical luminosity $L_c$ to get a
reversal from damping to excitation is the same for the eccentricity
and inclination. It is also the same as the critical luminosity at
which the thermal torque on a planet on a circular orbit
\citep{2017MNRAS.472.4204M} reverses from negative \citep[dominated by
the cold thermal torque, see][]{2014MNRAS.440..683L} to positive
\citep[dominated by the heating torque, see][]{2015Natur.520...63B},
and the same as well at which the thermal force on a luminous
perturber moving across a uniform medium reverses from drag to
thrust \citep{2019MNRAS.483.4383V}. While the thermal force on planets
in discs has not been studied numerically in a systematic fashion, the
recent numerical simulations of \citet{2019MNRAS.483.4383V}
corroborate with a high accuracy the value of the critical luminosity
in unsheared, homogeneous media.

The numerical simulations of \citet{2014MNRAS.440..683L},
\citet{2015Natur.520...63B} and \citet{2017arXiv170401931E} all seem
to indicate that the effects of thermal forces are cut off above a few 
Earth masses, for the set of parameters considered in these
studies. However, this effect has not been studied in detail and
should probably be tackled through numerical simulations.

The effects that we present here should have important consequences on
various stages of planet formation, such as the phase of oligarchic
growth or the giant impact stage, when those occur in the gaseous
disc. It could have also consequences on the trapping in mean motion
resonances of Earth-sized protoplanets. A detailed study of such
effects requires to track the thermal and luminous history of embryos,
taking into account the accretion of solid bodies and possibly the
effect of mutual collisions.

\section{Acknowledgments}

For this work, SF was supported by the Programa de Apoyo a Proyectos de Investigacion e Innovacion Tecnologica (PAPIIT) No IA101619  and FM was supported by the Programa de Apoyo a Proyectos de Investigacion e Innovacion Tecnologica (PAPIIT) No IN101616.

\appendix

\section{Impact of corotation offset}
\label{sec:dist-from-guid}
Our derivation assumes that the distance $x_p^0$ between corotation
and the epicycle's guiding centre is small compared to the radial and
vertical excursion of the planet. We hereafter discuss what happens when
this assumption is relaxed. It is convenient, in this case, to have
the origin of the $x$ axis at the planet's guiding centre, so that the
potential terms of Eqs.~\eqref{eq:118}, \eqref{eq:121}
and~\eqref{eq:122} keep the same form, whereas the unperturbed
azimuthal velocity has now the form:
\begin{equation}
  \label{eq:194} 
 v_0=-\frac 32\Omega_p\left(x+x_p^0\right).
\end{equation}
With a finite value of $x_p^0$, the relationships of
Eqs.~\eqref{eq:74} and~\eqref{eq:76} are no longer verified, as they
relied on the symmetry in $x$ of the response. The expressions of
the force components worked out in sections~\ref{sec:radial-force}
and~\ref{sec:azim-comp-force} can be expressed in a slightly more
general fashion as follows:
\begin{equation}
  \label{eq:195}
  F_x^S=\frac{E}{4\pi^2}\iiint\partial_x(\tilde\Phi_e^+-\tilde\Phi_e^-)\Im(\tilde\rho_c)-\partial_x\tilde\Phi_c\Im(\tilde\rho_e^+-\tilde\rho_e^-)dx\,d^2\bs k 
\end{equation}
and 
\begin{equation}
  \label{eq:196}
  F_y^C=\frac{E}{4\pi^2}\iiint 
  k_y\Im(\tilde\rho_c)(\tilde\Phi_e^++\tilde\Phi_e^-)+k_y\Im(\tilde\rho_e^++\tilde\rho_e^-)\tilde\Phi_c 
  dx\,d^2\bs k,
\end{equation}
where we have used the relations
\begin{equation}
  \label{eq:197}
\tilde\rho_c(x,-k_y,-k_z)=\tilde\rho_c(x,k_y,k_z)^*
\end{equation}
and
\begin{equation}
  \label{eq:198}
  \tilde\rho_e^+(x,-k_y,-k_z)=\tilde\rho_e^-(x,-k_y,-k_z)^*.
\end{equation}
Using Eqs.~\eqref{eq:118}, \eqref{eq:121} and~\eqref{eq:122} we write
\begin{equation}
  \label{eq:199}
  \tilde\Phi_e^+-\tilde\Phi_e^-=-2k_y\tilde\Phi_c 
\end{equation}
and 
\begin{equation}
  \label{eq:200}
  \tilde\Phi_e^++\tilde\Phi_e^-=\partial_x\tilde\Phi_c,
\end{equation}
to get rid of all instances of $\tilde\rho_c$ and $\tilde\Phi_c$ in
the time derivative of the eccentricity given by Eq.~\eqref{eq:178}, and
eventually obtain:
\begin{equation}
  \label{eq:201}
  \frac{\bar{\dot e}}{e}=\frac{1}{4\pi^2M\Omega_p}\iiint
 \left[ -\Im(\tilde\rho_e^+)\tilde\Phi_e^++\Im(\tilde\rho_e^-)\tilde\Phi_e^-\right]
  \,dx\,d^2\bs k.
\end{equation}
This, with our notation, is equivalent to the original derivation of
\citet{2004ApJ...602..388T} who checked that their force expression
was compatible with the time derivative of the eccentricity given by a
sum on the first-order Lindblad resonances
\citep{gt80,arty93b}. Although the expressions of the ``circular''
terms $\tilde\rho_c$ and $\tilde\Phi_c$ are required to obtain the
individual force components of Eqs.~\eqref{eq:172} to~\eqref{eq:175},
they cancel out in the expression of the time derivative of the
eccentricity. Assessing how the latter varies with $x_p^0$ is
therefore tantamount to assessing how $\tilde\rho_e^\pm$ varies with
$x_p^0$. Eq.~\eqref{eq:80} becomes:
\begin{equation}
  \label{eq:202}
  j(X_c-X_p^0-X)\tilde\rho_e^+=K(\tilde\rho_e^{+ ''}-\tilde\rho_e^+) + s_1\left[\delta(X)-\frac34X_c\delta'(X)\right], 
\end{equation}
where
\begin{equation}
  \label{eq:203}
  X_p^0\equiv x_p^0k.
\end{equation}
Calling $\tilde\rho_{e,0}^+$ the solution for $X_p^0=0$ and writing
$\delta\tilde\rho_e^+=\tilde\rho_e^+-\tilde\rho_{e,0}^+$, we have:
\begin{equation}
  \label{eq:204}
  j(X_c-X_p^0-X)\delta\tilde\rho_e^+=K(\delta\tilde\rho_e^{+"}-\delta\tilde\rho_e^+)+iX_P^0\tilde\rho_{e,0}^+.
\end{equation}
From Fig.~\ref{fig:forces}, it is evident that the spatial frequencies
that most contribute to the response are $|K_y|\sim |K_z|\sim 1$
(hence $k_{y,z}\sim \lambda_c^{-1}$), and for those $|X_c|\sim 1$ and
$X_p^0\sim x_p^0/\lambda_c$. From Eq.~\eqref{eq:204}, we deduce that
as long as $|X_p^0|\ll|X_c|$ (i.e. $x_p^0\ll \lambda_c$), the
correction $\delta\tilde\rho_e^+$ is small compared to the symmetric
solution $\tilde\rho_{e,0}^+$.  The time derivative of the
eccentricity is therefore correct as long as $|x_p^0|\ll \lambda_c$,
regardless of whether it is smaller or larger than the epicyclic
excursion $E$, and the growth of eccentricity when $L>L_c$ is not a
finite amplitude instability. The numerical experiments of
\citet[][e.g. their Fig.~14]{2017arXiv170401931E} agree with this
statement.  Note that similar arguments apply to the damping of
eccentricity arising from wave launching.  The expression of
\citet{2004ApJ...602..388T} should be valid all the way to
eccentricities much smaller than the offset between corotation and
guiding centre. On physical grounds, the eccentricity varies because
the perturber is subjected to a force that depends on its position on
the epicycle. As long as the corotation offset is not too large,
shifting corotation amounts to adding a constant force which has no
impact on the eccentricity budget.

Strictly similar considerations apply to the excitation or damping of
the inclination.

\section{Extension to a viscous disc}
\label{sec:extens-visc-disc}
Our derivation has not considered how viscous heating would affect the
energy budget in Eq.~\eqref{eq:5}.  If we consider a laminar disc with
kinematic viscosity $\nu$, the dominant source term for the viscous
heating is:
\begin{equation}
  \label{eq:205}
  S_d(\bs r) = \frac 12\rho\nu(\partial_xv)^2
\end{equation}
and the dominant term arising from the perturbation induced by the
planet is, from Eq.~\eqref{eq:11}:
\begin{equation}
  \label{eq:206}
  S'_d(\bs r)=\frac 98\nu\Omega_p^2\rho'-3\rho_0\nu\Omega_p\partial_xv',
\end{equation}
where, for lengthscales typical of that of the perturbation,
$|\partial_xv'|\sim |v'|/\lambda$.  Eq.~\eqref{eq:30}--\eqref{eq:32}
imply that, in order of magnitude for $|x|\sim\lambda$,
$v'\sim GM/(R^2\Omega_p)$, where $R$ is the distance to the
planet. We can then write the order-of-magnitude relationships
$|v'|\sim (R_B/R)H^2\Omega_p/R\sim(\rho'/\rho_0)H^2\Omega_p/\lambda$,
where $R_B=GM/c_s^2$ is the planetary Bondi
radius. Eq.~\eqref{eq:206} is then dominated by its second term, which
has order of magnitude $\rho'\nu\Omega_p^2(H/\lambda)^2$. This source
term has to be compared to the divergence of the heat flux, which has
the order of magnitude
$\chi c_s^2\rho'/\lambda^2\sim\rho'\chi\Omega_p^2(H/\lambda)^2$. In
discs that have a Prandtl number $\mathrm{Pr}\equiv\chi/\nu \gg 1$, the viscous
dissipation induced by the planetary perturbation is negligible
compared to the heat flux. Our analysis should remain valid in such
discs. The numerical exploration of \citet{2017arXiv170401931E} took
place in a disc with $\mathrm{Pr}\sim 5$ and yielded results
compatible with this expectation.



\bibliographystyle{mnras}

\bsp
\label{lastpage}
\end{document}